\documentclass{JHEP3}
\usepackage{amssymb}
\usepackage{amsmath}
\usepackage{amstext}
\usepackage{mcite}

\usepackage[pdftex]{graphicx}

\def\a{\alpha}
\def\b{\beta}
\def\g{\gamma}
\def\d{\delta}

\def\l{\lambda}
\def\m{\mu}
\def\n{\nu}

\def\p{\pi}
\def\r{\rho}
\def\s{\sigma}
\def\t{\tau}

\def\G{\Gamma}
\def\D{\Delta}
\def\L{\Lambda}

\def\W{\Omega}

\newcommand{\eq}[1]{(\ref{#1})} 
\newcommand{\pd}[2]{\frac{\partial #1}{\partial #2}} 

\title{Holographic Roberge-Weiss Transitions II: Defect Theories and the Sakai-Sugimoto Model.}
\author{James Rafferty\\
Department of Physics, Swansea University, Singleton Park, Swansea, SA2 8PP.\\
email: \email{pyjames@swansea.ac.uk} }
\abstract{We extend the work of \cite{Aarts:2010ky}, including an imaginary chemical potential for quark number into the Sakai-Sugimoto model and codimension $k$ defect theories. The phase diagram of these models are a function of three parameters, the temperature, chemical potential and the asymptotic separation of the flavour branes, related to a mass for the quarks in the boundary theories. We compute the phase diagrams and the pressure due to the flavours of the theories as a function of these parameters and show that there are Roberge-Weiss transitions in the high temperature phases, chiral symmetry restored for the Sakai-Sugimoto model and deconfined for the defect models, while at low temperatures there are no Roberge-Weiss transitions. In all the models we consider the transitions between low and high temperature phases are first order, hence the points where they meet the Roberge-Weiss lines are triple points. The pressure for the defect theories scales in the way we expect from dimensional analysis while the Sakai-Sugimoto model exhibits unusual scaling. We show that the models we consider are all analytic in $\m^2$ when $\m^2$ is small.}
\preprint{}
\begin{document}
\section{Introduction}
Since the original discovery by Maldacena of a duality between type IIB supergravity on AdS$_5 \times$ S$^5$ and $\mathcal{N}=4$ super Yang-Mills at strong coupling \cite{Maldacena:1997re} there has been a consistent effort to find a holographic model that is more like the theories we observe in nature, i.e. contain less supersymmetry and the correct degrees of freedom. In particular, much research has been dedicated to finding QCD like models at strong coupling using a classical or semiclassical supergravity. One such model that has seen some success is that of Sakai and Sugimoto \cite{Sakai:2004cn} which is a holographic model in type IIA supergravity, dual to a QCD like large $N$ theory, where $N$ is the number of colours. The boundary geometry is $\mathbb{M}^4 \times S^1$ so at suitable energy scales the gauge theory is effectively 4 dimensional, furthermore applying antiperiodic boundary conditions for fermions on the $S^1$ breaks supersymmetry so the only remaining degrees of freedom are those of large $N$ QCD. Since its discovery, much progress has been made in understanding how quark number \cite{Horigome:2006xu} and isospin \cite{Aharony:2007uu,Parnachev:2007bc} chemical potentials affect the phases of the model, and also the effect of chemical potentials in holographic models in general~\cite{Ishihara:2008zzb,*Mateos:2007vc,*Mateos:2007vn,*Karch:2007br,*Gubser:1998jb,*Chamblin:1999tk,*Cvetic:1999ne,*Babington:2003vm,*Kirsch:2004km,*Albash:2006ew,*Karch:2006bv, *Bigazzi:2011it, Yee:2009cd}. 

Another approach to studying field theories is to analyse them in different dimensions. It has been known for many years that the gauge theories in different dimensionality to the usual $3+1$ we are familiar with can behave in very different ways than expected. Furthermore, much effort has gone into studying field theories in $1+1$ dimensions, as a much more general class of theories can be solved exactly than in 4 dimensions. One may produce gauge theories in AdS/CFT with different dimensionality by altering the dimensionality of the branes that go into producing the background geometry. In particular, we will be interested in systems where the colour sector lives in $3+1$ dimensions but the flavour sector lives on a $3-k+1$ dimensional submanifold of the boundary, dual to a background of $N$ D3 branes with $N_f$ D$(7-2k)$ branes in the probe approximation added. The intersection of these branes forms a \textit{defect}. It was found in \cite{Benincasa:2009be} that the order of the confinement/deconfinement phase transition in the $T / \m$ plane depends on the codimension of the defect for real chemical potentials.

The motivation for studying theories with an imaginary chemical potential comes from the sign problem in lattice gauge theory, which makes simulating QCD with a real chemical potential hard to accomplish because the fermion determinant for such a system is complex. A way to circumvent this problem using standard lattice techniques is to take the chemical potential to be a purely imaginary number, and systems where this is the case have been extensively studied in lattice gauge theory \cite{deForcrand:2010ys,*deForcrand:26010he,*deForcrand:2006pv,*deForcrand:2003hx,*deForcrand:2002ci,*D'Elia:2007ke,*D'Elia:2004at,*D'Elia:2002gd,*D'Elia:2009qz,*D'Elia:2009tm}. 

The goal of this paper is to compute the phase diagram for the Sakai-Sugimoto model and for codimension $k=1$ and $k=2$ defect theories in the presence of an imaginary chemical potential. In section \ref{sec:rw} we provide a short review of the analysis of Roberge and Weiss \cite{Roberge:1986mm} who studied a field theory system with an imaginary chemical potential in perturbation theory. In section \ref{sec:d3d7} we describe the results of \cite{Aarts:2010ky} where the D3 - D7 system, dual to $\mathcal{N}=4$ supersymmetric Yang-Mills with $\mathcal{N}=2$ fundamental matter at strong coupling with an imaginary chemical potential was studied. Sections \ref{sec:SS} and \ref{sec:defect} give a short review of and the Sakai-Sugimoto model at low and high temperatures \cite{Sakai:2004cn,Aharony:2006da,Parnachev:2006dn} and codimension $k$ defect theories respectively followed by the computation of their phase diagrams. We conclude with a brief discussion and future outlook.
\section{Roberge-Weiss Transitions}
\label{sec:rw}
In 1986 Roberge and Weiss studied QCD with an imaginary chemical potential for quark number \cite{Roberge:1986mm}. It is well understood that the phases of a pure $SU(N)$ gauge theory at finite temperature can be distinguished by the expectation value of the Polyakov loop $\langle P \rangle$. At low temperatures the theory is in the confined phase which implies the expectation value of the phase of the Polyakov loop vanishes. This further implies there is a $\mathbb{Z}_N$ symmetry where the Polyakov loop transforms as $P \to e^{\frac{2 \p i r}{N}}P$ with $r$ an integer. For high temperatures this symmetry is spontaneously broken and hence $\langle P \rangle \neq 0$ so the theory becomes deconfined. When one introduces fundamental matter into the theory the $\mathbb{Z}_N$ symmetry is never a symmetry of the theory, however a remnant of the $\mathbb{Z}_N$ symmetry is still present. Introducing an imaginary chemical potential makes this explicit and resolves the ambiguity between confined and deconfined phases of the theory. Consider the partition function for such a theory
\begin{equation}
Z\left[\m_I\right] = \mathrm{Tr} \left(e^{-\b H + i \b \m_I N_q}\right)
\end{equation}
$N_q$ is the quark number operator and $\m_I$ is the imaginary chemical potential\footnote{Note that we have made the replacement $\m \to i \m_I$ so $\m_I$ on it's own is in fact real.}. One can see this partition function has a symmetry under the transformation $\m_I \to \m_I + \frac{2\p}{\b}$ where $\b$ is the inverse temperature of the system. Together with the fact that $N_q$ is quantised, this implies $\b \m_I \in \left[-\p ,\p\right]$.

The adjoint sector also contains a $\mathbb{Z}_N$ symmetry which introduces a further periodicity into the theory. The imaginary chemical potential couples to fundamental matter in the same way as the time component of the $SU(N)$ gauge field. Hence, the time component of the gauge covariant derivative associated to this gauge field is modified by the presence of this chemical potential in the following way
\begin{equation}
\partial_\t + i A_\t \to \partial_\t + i A_\t -i \m_I
\end{equation}
We can remove the dependence of the action on $\m_I$ by rescaling the fields so that the dependence on $\m_I$ is in the boundary conditions around the thermal circle
\begin{equation}
\phi \left(\vec{x},\b\right) \sim e^{i \m_I \b} \phi \left(\vec{x},0\right)
\end{equation}
where the fermionic and bosonic modes differ by a sign. We may now apply an $SU(N)$ gauge transformation with the group element $U\left(\vec{x},\t\right)$ with the property that $U\left( \vec{x},\b\right) = e^{\frac{2 \p i r}{N}}U\left( \vec{x},0\right)$ and $r$ an integer. The action and path integral measure are left invariant but the boundary conditions around the thermal circle are not, hence
\begin{equation}
\phi \left(\vec{x},\b\right) \sim e^{i \left( \m_I \b + \frac{2 \p r}{N} \right) } \phi \left(\vec{x},0\right)
\end{equation}
The partition function must be invariant under these transformations, therefore
\begin{equation}
Z\left[\m_I\right]= Z\left[\m_I + \frac{2 \p r }{\b N} \right]
\end{equation}
which further implies $\b \m_I \in \left[-\frac{\p}{N} ,\frac{\p}{N}\right]$.

The free energy $F\left[\m_I\right] = -T \log \left[Z\right]$ of QCD in perturbation theory, valid at high temperature, was first computed in \cite{Roberge:1986mm} who found first order phase transitions at constant values of $\m_I$. These values are
\begin{equation}
\m_\mathrm{RW} = \frac{\left(2r-1\right) \p T}{N_c} \label{eq:murw}
\end{equation}
Lattice studies suggest that at low temperature when the theory is strongly coupled there are no phase transitions and the physics is smooth as a function of $\m_I$, while at high temperature there are first order Roberge-Weiss transitions at $\m_I = \m_\mathrm{RW}$ \cite{deForcrand:2010ys,*deForcrand:26010he,*deForcrand:2006pv,*deForcrand:2003hx,*deForcrand:2002ci,*D'Elia:2007ke,*D'Elia:2004at,*D'Elia:2002gd,*D'Elia:2009qz,*D'Elia:2009tm}.
\section{The D3-D7 System}
\label{sec:d3d7}
It is possible to study $\mathcal{N}=4$ super Yang-Mills with $\mathcal{N}=2$ fundamental matter at strong coupling with an imaginary chemical potential analytically via the celebrated AdS/CFT correspondence. In order to introduce a chemical potential into the D3 - D7 model it is necessary to introduce a non zero background time component of a worldvolume gauge field on the D7 branes. Furthermore in the Euclidean formulation to get an imaginary chemical potential this constant value of the gauge field must be real. The quark number chemical potential $\m$ of the boundary gauge theory is given holographically by
\begin{equation}
\m = \lim_{r \to \infty} A_t(r)
\end{equation}
$A_t(r)$ is the $t$ component of a $U(1)$ gauge field and $r$ is a radial coordinate where $r \to \infty$ is the boundary limit. The gauge field strength is $F = F_{\m \n} dx^\m \wedge dx^\n$ and since we do not wish to study the theory as a function of a conductivity the only non zero component is $F_{t r} dr \wedge dt$. Since we work in Euclidean signature one can see that the gauge field must be completely real in order that the the chemical potential is purely imaginary since when we Wick rotate the time direction the field strength becomes $i F_{\t r} dr \wedge d\t$.
\FIGURE{
\centering
\includegraphics[width=13cm]{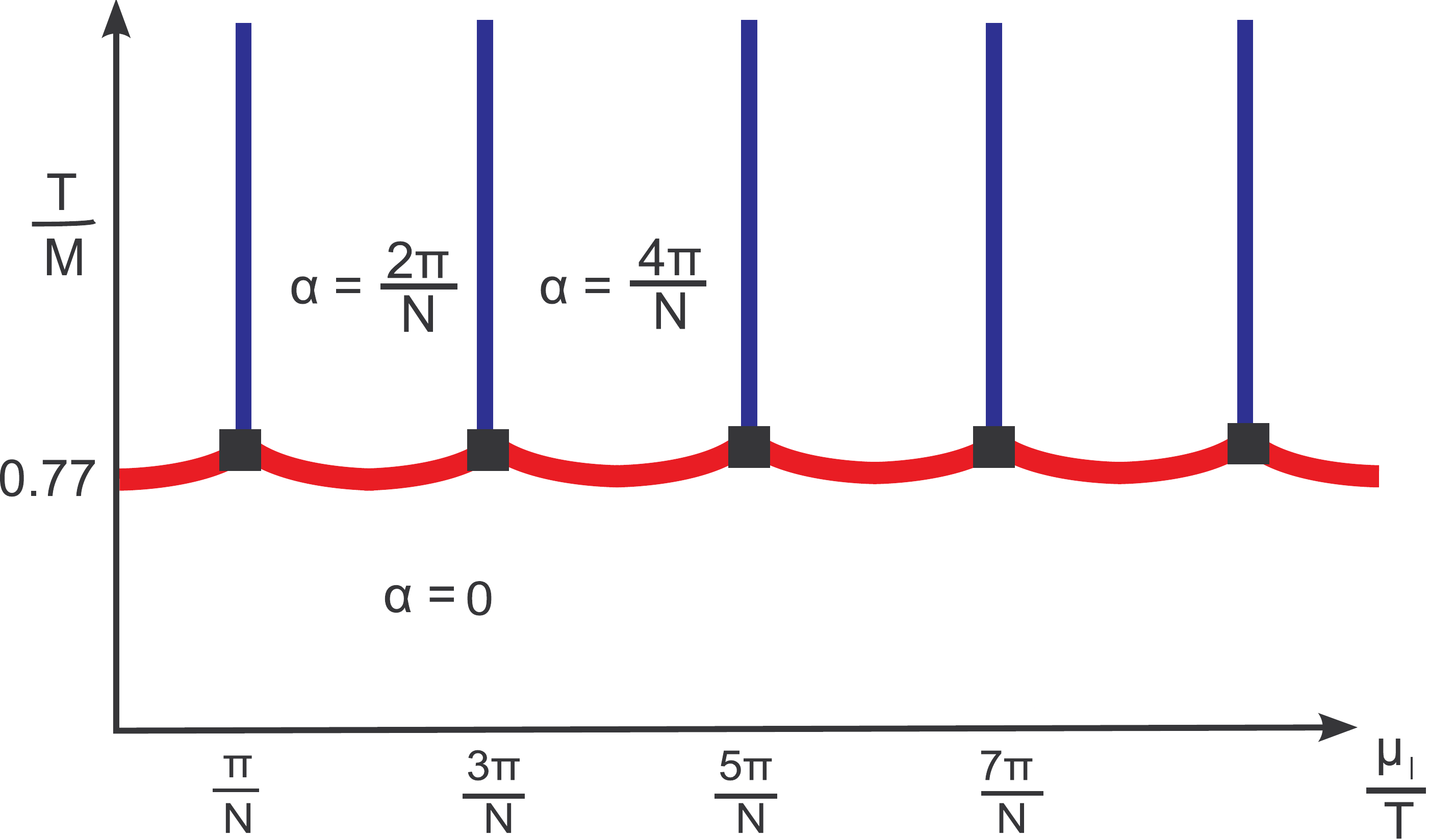}
\caption{Phase diagram of the D3 - D7 system with an imaginary chemical potential.}
\label{fig:phased3}
}
The features of the phase diagram are as follows; At low temperatures there are no phase transitions for varying chemical potential. The first order meson melting transition that appears for zero chemical potential extends into a first order line at $T = T_c\left(\m_I\right)$. The dependence of $T_c$ on $\m_I$ was found to be
\begin{equation}
T_c(\m_I) = T_0 + K \frac{m}{\l} \left(\frac{\m_I}{T_0} - \a \right)^2 + \ldots 
\end{equation}
Where $T_0 = T_c(0) \approx 0.77$, $K \approx 33.5$, $\l$ is the 't Hooft coupling and $m$ is the quark mass. $\a$ is the expectation value of the phase of the Polyakov loop. Higher corrections are suppressed by factors of $\frac{1}{N}$ since $\left| \frac{\m_I}{T_c(0)} - \a\right| < \frac{\p}{N}$. Above this temperature there is an infinite set of Roberge-Weiss transitions at constant values of $\m_I = \m_{\mathrm{RW}}=\frac{\left(2r-1\right) \p T}{N}$.\footnote{One might imagine that, since the spacing of the Roberge-Weiss phases is of order $1/N$, we should consider a model that goes beyond the supergravity limit, however, the factors of $N$ arise from the symmetry of the theory under shifts of $\m_I \to \m_I + \frac{2 \p}{\b N}$ so it is consistent to consider the supergravity limit only.} These Roberge-Weiss lines are first order and so the points where they meet the meson melting line are triple points. 

The argument of \cite{Aharony:1998qu} (described in a later section) leading to the presence of Roberge-Weiss transitions in gravity duals is quite general. We expect a similar structure in both the defect theories we consider and the Sakai-Sugimoto model from the argument given in \cite{Yee:2009cd}. The details of the theory such as the $T_0 = T_c(\m_I = 0 )$ and the phase transition that takes the place of the meson melting transition here will not be the same however.
\section{The Sakai-Sugimoto Model}
\label{sec:SS}
In order to model QCD at strong coupling holographically the Sakai-Sugimoto model takes a stack of $N$ D4 colour branes in type IIA supergravity and in the large $N$ limit replaces them with a background geometry. To incorporate $N_f$ flavours D8 - $\overline{\mathrm{D}8}$ branes are added which, if $N_f \ll N$ we can consider in the probe approximation. To reduce the dimensionality of the boundary from 5 to 4 one of the dimensions is compactified on a circle, so if we consider meson masses below the Kaluza Klein mass scale $M_\mathrm{KK}$ the boundary gauge theory is effectively 4 dimensional. In addition, imposing antiperiodic boundary conditions for fermions on the compact direction breaks supersymmetry completely making the only light degrees of freedom those of large $N$ QCD. A great successes of the Sakai-Sugimoto model is it is a holographic model that realises chiral symmetry breaking. The geometry resulting from the D4 branes contains a horizon, and the D8 - $\overline{\mathrm{D}8}$ solutions that end at some finite height above that horizon represent a state in the field theory where chiral symmetry is broken. If one increases the temperature with the other parameters of the theory fixed the distance from the horizon to the end point of the D8 - $\overline{\mathrm{D}8}$ branes is reduced. Eventually, the horizon will meet the branes and chiral symmetry is restored. It has been shown in \cite{Aharony:2006da} that there are two bulk geometries with similar asymptotics, corresponding to QCD at low and high temperatures. For low temperature
\begin{equation}
ds^2= \left(\frac{U}{R}\right)^{\frac{3}{2}} \left(dt_E^2 + \d_{ij} dx^i dx^i + f_{KK}(U) d x_4^2 \right) + \left(\frac{R}{U}\right)^{\frac{3}{2}} \left( \frac{dU}{f_{KK}(U)} + U^2 d\W_4^2\right)\label{eq:coldmetric}
\end{equation}
with $f_{KK}(U) = 1-\left(\frac{U_{\mathrm{KK}}}{U}\right)^3$. $R$ is the radius of curvature and is related to the string coupling and string  length. $U_{\mathrm{KK}}$ is the position of the horizon and both are related to the Kaluza Klein mass $M_{\mathrm{KK}} = \frac{3}{2} \sqrt{\frac{U_{\mathrm{KK}}}{R^3}}$, which is the mass that characterises the scale of the compact direction. For the high temperature deconfined phase we have
\begin{equation}
ds^2= \left(\frac{U}{R}\right)^{\frac{3}{2}} \left(f_T(U) dt_E^2+ \d_{ij} dx^i dx^j +  d x_4^2 \right) + \left(\frac{R}{U}\right)^{\frac{3}{2}} \left( \frac{dU}{f_T(U)} + U^2 d\W_4^2\right) \label{eq:metric}
\end{equation} 
where $f_T(U)= 1-\left(\frac{U_{T}}{U}\right)^3$. Note we have Wick rotated to Euclidean signature, and to avoid a conical singularity at the horizon the Euclidean time direction must be periodically identified, hence there are two compact directions in the geometries and thus one can think of the low / high temperature behaviour as a competition between which circle shrinks to zero size first. Ensuring regularity at the horizon one can deduce that the temperature is given by
\begin{equation}
T = \frac{3}{4 \p} \sqrt{\frac{U_T}{R^3}} \label{eq:T}
\end{equation}
We will consider both background geometries, but we will pay particular attention to the high temperature phase because we expect to see Roberge-Weiss transitions there. In this case the $t_E$ circle shrinks to zero size, and thus the spacetime ends, at $U = U_T$ while the $x_4$ circle remains finite sized there. 

The phase diagram of the Sakai-Sugimoto model will be a function of 3 variables, the temperature and chemical potential are obvious but there is also another, less evident parameter. In the gravity picture there is a parameter $L$ that is the asymptotic separation of the D8 - $\overline{\mathrm{D}8}$  branes in the $x_4$ direction. Since in the effective field theory this direction is compactified and integrated out, operators like $\psi \g^\m A_\m \overline{\psi}\psi \g^\n A_\n \overline{\psi}$ will be generated with coefficients that depend on $L$. These operators are irrelevant, so it is not clear exactly how to relate $L$ to the various generated operators when the coupling is large. We will show below that the temperature and chemical potential naturally combine with $L$ to form dimensionless parameters, which we will use to plot the phase diagram.
\subsection{D8 Embedding}
Consider the metric for the D8 branes in the background \eq{eq:metric}. We choose the brane to share all coordinates except the $x_4$ direction where the D8 is pointlike, which we make a function of the radial direction $U$. The induced metric of the D8 with this choice is
\begin{align}
ds_{D8}^2 = \left(\frac{U}{R}\right)^{3/2}\left(f_T(U) dt_E^2+ \d_{ij} dx^i dx^j \right) +\left( \left(\frac{R}{U}\right)^{\frac{3}{2}}\frac{1}{f_T(U)}+\left(\frac{U}{R}\right)^{\frac{3}{2}} x_4'^2\right) dU^2 + \nonumber \\ +\left( \frac{R}{U}\right)^{\frac{3}{2}} U^2 d\W^2_4 \label{eq:pullback}
\end{align}
Prime denotes differentiation with respect to $U$. To incorporate a chemical potential into the gauge theory, on the gravity side we include a worldvolume gauge field on the D8 branes. We also introduce a NS B field into the bulk, dual to the expectation value of the phase of the Polyakov loop in the gauge theory. The only non-zero components of these field strengths are $F_{tU}$ and $B_{tU}$. We can now write the DBI action of $N_f$ branes
\begin{align}
S_{\mathrm{DBI}} &= N_f T_{\mathrm{D}8} \int d^8 x e^{-\phi} \sqrt{\det\left(g_{ab} + B_{ab} + 2 \p \a' F_{ab} \right) }\\
&= \frac{N_f T_{\mathrm{D}8} \mathrm{Vol}(S^4)}{g_s}\int d^4 x dU U^{4} \sqrt{\frac{R^3}{U^3} \left(1 +\left(B_{tU}+ 2 \p \a' F_{tU}\right)^2\right)+ f_T (U) x_4'^2} \label{eq:ssaction}
\end{align}
$T_{\mathrm{D}8}$ is the brane tension and we have used that the dilaton is given by
\begin{equation}
e^\phi = g_s \left(\frac{U}{R} \right)^{\frac{3}{4}}
\end{equation}
Furthermore, we may set the $U$ components of the gauge potentials to zero with an appropriate choice of gauge, therefore $F_{tU} = -\partial_U A_t = -A'$ hence Chern Simons terms do not contribute to the action. 

Since the action only depends on derivatives of $x_4$ and $A$ we have very simple equations of motion that after one integration take the following form; The equation of motion for $x_4$ is given by
\begin{equation}
\frac{U^4 f_T(U) x_4'}{\sqrt{f_T(U) x_4'^2 + \left(\frac{R}{U} \right)^3 \left(1 +\left(B_{tU}+ 2 \p \a' F_{tU}\right)^2  \right)}} = \frac{ k g_s}{N_f   T_\mathrm{D8} \text{Vol}\left(S^4\right) } = \tilde{k} \label{eq:eom}
\end{equation}
and the equation of motion for $F_{tU}$ is
\begin{equation}
\frac{U R^3 \left(B_{tU}+ 2 \p \a' F_{tU}\right)}{\sqrt{f_T(U) x_4'^2 + \left(\frac{R}{U} \right)^3 \left(1 +\left(B_{tU}+ 2 \p \a'  F_{tU}\right)^2 \right)}} = \frac{ d g_s}{2 \p \a' N_f   T_\mathrm{D8} \text{Vol}\left(S^4\right) } = \tilde{d} \label{eq:eq2} 
\end{equation}
where $k$ and $d$ are constants determined by boundary conditions. There are two different types of brane profile allowed by these equations of motion. The first is where the brane extends all the way to the thermal horizon, covering all of the radial direction, and secondly, where the brane turns around meeting a $\overline{\mathrm{D}8}$ at some finite radius $U_0$. In the low temperature phase only the latter solution is allowed, because the $x_4$ circle in the low temperature geometry shrinks to zero size first while the thermal circle remains finite sized over the whole range of the radial coordinate. The converse is true in the high temperature case, allowing both types of solution described above. These solutions are dual to the chiral symmetric and broken phases in the boundary theory respectively. 

It is straightforward to fix the constants above by considering boundary conditions. Consider \eq{eq:eom} at some point such that $x_4'\left(U\right) = 0$. We find
\begin{equation}
\tilde{k}=\mathcal{O} \left(x_4'\right)
\end{equation}
Therefore $\tilde{k}$ is forced to vanish. We will find a useful constraint on $\tilde{d}$ for this type of embedding later.

Consider now the other type of solution where the brane turns around. There must be some point $U = U_0$ where the slope of $x_4(U)$ diverges so consider $x_4'(U = U_0) \to \infty$. From \eq{eq:eom} we find
\begin{equation}
\tilde{k}=U_0^{4} \sqrt{f_T(U_0)}
\end{equation}
and from equation \eq{eq:eq2}
\begin{equation}
\tilde{d}= \mathcal{O} \left(\frac{1}{x_4'}\right) 
\end{equation}
So for this type of embedding, $\tilde{d}$ is forced to vanish\footnote{This is strictly $\tilde{d}= \mathcal{O} \left(\frac{F_{Ut}}{x_4'}\right)$ which does not necessarily constrain $\tilde{d}$ to vanish if $F_{Ut}$ diverges. One may assume $U=U(x_4)$ rather than $x_4 = x_4(U)$ as used to derive \eq{eq:pullback} to show that this is not the case and $\tilde{d}$ does indeed vanish for this type of embedding.}. Note: This analysis is analogous to the real chemical potential case studied in \cite{Horigome:2006xu}.

Consider the case where the D8 branes extend all the way to the thermal horizon; from equation \eq{eq:eq2} we can solve for $\left(B_{tU}+2 \p \a' F_{tU}\right)$ and plug back into \eq{eq:eom}. Doing this we find
\begin{equation}
\left(B_{tU}+2 \p \a' F_{tU}\right)^2 = \frac{\tilde{d}^2\left(1+ \frac{U^3}{R^3} f_T(U) x_4'^2 \right) }{ R^3 U^5-\tilde{d}^2 } \label{eq:bplusf}
\end{equation}
and hence
\begin{equation}
\frac{U^4 f_T(U) x_4'}{\sqrt{f_T(U) x_4'^2 + \left(\frac{R}{U} \right)^3 \left(1 +\frac{\tilde{d}^2\left(1 + \left(\frac{U}{R} \right)^3 f_T(U) x_4'^2\right) }{  U^5 R^3 - \tilde{d}^2 }  \right)}}  = 0\label{eq:eom2}
\end{equation}
One can see that \eq{eq:eom2} admits constant solutions that cover the entire range of the $U$ direction, falling into the horizon at $U= U_T$. Since for these solutions $x_4$ is arbitrary and any value of $x_4$ will be have the same action, these solutions must correspond to the phase where chiral symmetry is unbroken and the quarks in the boundary theory are exactly massless. To ensure the reality of $\left(B_{tU}+2 \p \a'F_{tU}\right)$ to produce an imaginary chemical potential in the gauge theory, and to ensure the reality of $x_4$ one can see there is a constraint on $\tilde{d}$ from \eq{eq:bplusf}
\begin{equation}
\tilde{d}^2 \leq  R^3 U_T^5 \label{eq:range}
\end{equation}
In contrast to the condition derived in \cite{Aarts:2010ky}, the condition here does not depend on the horizon value of $x_4(U)$. In the D3/D7 case the bound on $\tilde{d}$ depended on the asymptotic value of the slipping mode, or the quark mass in the language of the gauge theory. As described above this type of solution only corresponds to strictly massless quarks.
\subsection{The Potential}
\label{sec:pressure}
\subsubsection{The unbroken chiral symmetry case}
Firstly, we note that for all phases of the Sakai-Sugimoto model the Lagrangian quark mass is zero, furthermore in this phase there can be no dynamically generated quark mass as that would break chiral symmetry. As mentioned previously it was shown in \cite{Aarts:2010ky}, the chemical potential is related to the worldvolume gauge field on the gravity side while the expectation value of the Polyakov loop in the theory is related to the NS-NS 2 form. Together we have
\begin{equation}
\a - \b \m_I = \frac{1}{2 \p \a'}\int_{D_2} \left(B_{tU}+2 \p \a'F_{tU}\right) \label{eq:aplusb}
\end{equation}
where the integral is over the thermal cigar. Plugging into this using \eq{eq:bplusf} we find an explicit expression for $\a - \b \m_I$
\begin{equation}
\a - \b \m_I = \frac{\b}{2 \p \a'} \int_{U_T}^\infty dU \frac{\tilde{d}}{\sqrt{U^5 R^3 -\tilde{d}^2}} \label{eq:mu}
\end{equation}
To find the potential we evaluate the (renormalised) action and eliminate $\tilde{d}$ to find $S$ in terms of $\a - \b \m_I$. In the simple case of black hole embeddings the integrals can be done analytically
\begin{equation}
\left(\a - \b \m_I\right) = \frac{d^{\frac{2}{5}}}{2 \pi \alpha' T } \left(\frac{(-1)^{\frac{3}{10}} \G\left(\frac{3}{10}\right) \G\left(\frac{6}{5}\right)}{  \sqrt{\p} R^{\frac{3}{5}}}-\frac{d^{\frac{3}{5}} U_T \sqrt{1 -\frac{  R^3 U_T^5}{d^2}} {}_2 F_1 \left(\frac{1}{5},\frac{1}{2},\frac{6}{5},\frac{R^3 U_T^5}{d^2}\right)}{\sqrt{R^3 U_T^5-d^2}}\right)
\end{equation}
Plugging into the action \eq{eq:ssaction} with the equations of motion and the constraints found for black hole embeddings above we find
\begin{equation}
S = \frac{\sqrt{R^3} N_f T_{\mathrm{D}8} \mathrm{Vol} \left(S^4\right) \mathrm{Vol} \left(\mathbb{R}^3 \right)\b}{g_s} \int_{U_T}^\infty dU \frac{U^5}{\sqrt{U^5 - \frac{\tilde{d}^2}{R^3}}}
\end{equation}
Let us relabel the overall constant $\mathcal{A} = \frac{\sqrt{R^3} N_f T_{\mathrm{D}8} \mathrm{Vol} \left(S^4\right) \mathrm{Vol} \left(\mathbb{R}^3 \right)\b}{g_s}$.  This integral is divergent, so after regulating we may add a counterterm to cancel the divergence
\begin{equation}
S =  \mathcal{A} \int_{U_T}^\L dU \frac{U^5}{\sqrt{U^5 - \frac{\tilde{d}^2}{R^3}}} - \mathcal{A} \frac{2 \L^{7/2}}{7}
\end{equation}
Evaluating the integral and taking the limit $\L \to \infty$ we find
\begin{equation}
S = \frac{2}{7} \mathcal{A} U_T \sqrt{U_T^5-\frac{\tilde{d}^2}{R^3}}\left(\frac{(-1)^{\frac{3}{10}} \tilde{d}^{\frac{7}{5}} \G\left(\frac{3}{10}\right) \G\left(\frac{6}{5}\right)}{ \sqrt{\pi } R^{\frac{21}{10}} U_T \sqrt{U_T^5-\frac{\tilde{d}^2}{R^3}} }-1 + \frac{{}_2 F_1 \left(\frac{1}{5},\frac{1}{2},\frac{6}{5},\frac{R^3 U_T^5}{\tilde{d}^2}\right)}{\sqrt{1-\frac{R^3 U_T^5}{\tilde{d}^2}}}\right)
\end{equation}
Using $\tilde{d}$ as a parameter we find figure \ref{fig:potential}.
\FIGURE{
\centering
\includegraphics{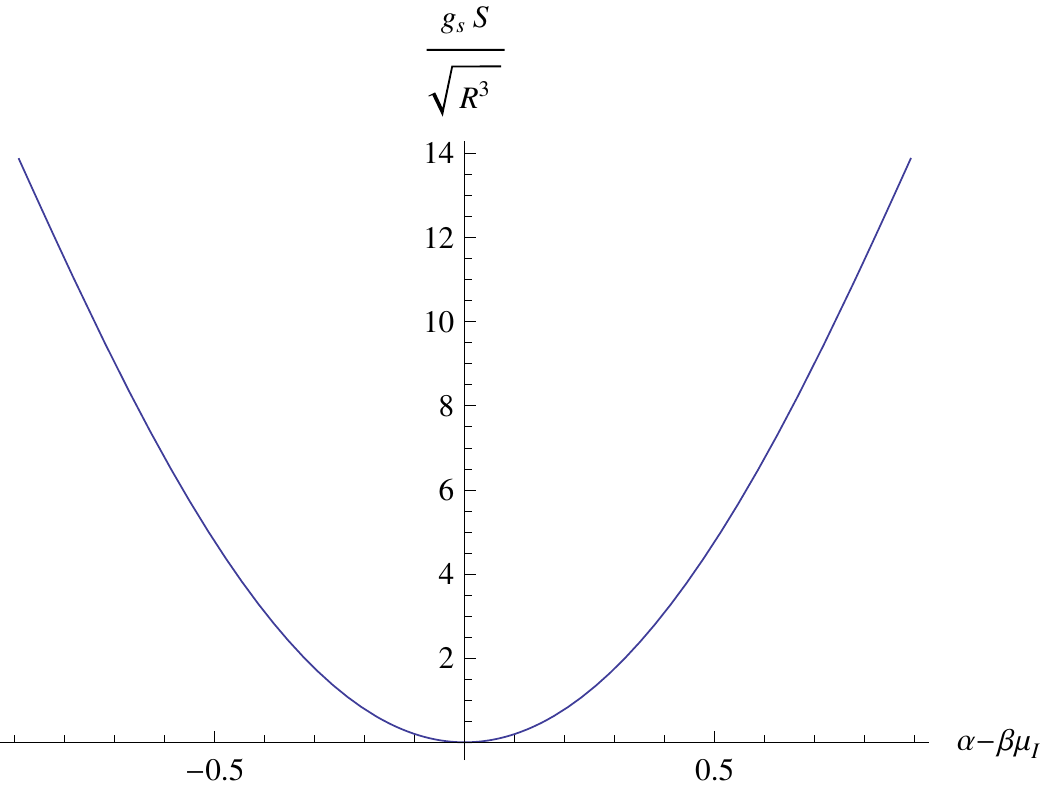}
\caption{The Roberge-Weiss potential as a function of $\a - \b \m_I$}
\label{fig:potential}
}
Since the potential is truncated at order $\frac{1}{N}$, it is instructive to expand these expressions for small $\tilde{d}$ and eliminate to find $S$ in terms of $\a - \b \m_I$, we find the first order correction to the action as a function of $(\a - \b \m_I)$
\begin{align}
\a - \b \m_I &\approx \frac{\tilde{d}}{3 \pi  \alpha'  T R^{\frac{3}{2}}  U_T^{\frac{3}{2}} } + \mathcal{O} \left(\tilde{d}^3 \right)\\
S &\supset \frac{\mathcal{A} \tilde{d}^2}{3 R^3 U_T^{\frac{3}{2}}} + \mathcal{O} \left(\tilde{d}^4 \right)
\end{align}
hence
\begin{equation}
S \supset 3  \pi ^2 \alpha'^2  \mathcal{A} T^2 U_T^{\frac{3}{2}}   (\a - \b \m_I) ^2 + \mathcal{O} \left((\a - \b \m_I) ^4 \right)
\end{equation}
We wish to write this expression in terms of field theory quantities; in the high temperature phase the $x_4$ direction is finite sized over the whole range of the radial coordinate so there is no constraint on it to ensure regularity at the horizon. The size of the $x_4$ direction is therefore related to the Kaluza Klein mass scale simply by
\begin{equation}
\frac{1}{M_\mathrm{KK}} =2 \p R_{x_4}
\end{equation}
The tension of the D8 brane is a dimension 9 object related to the string length by $T_{\mathrm{D}8}=\left( \left(2 \p\right)^8 l_s^9\right)^{-1}$. We can relate the radius of curvature and the string coupling to gauge theory quantities using the following (from \cite{Aharony:2006da})
\begin{equation}
\frac{R^3}{l_s^3} = \p g_s N, \quad g_s = \frac{g_4^2}{\left(2 \p\right)^2 l_s M_\mathrm{KK}}
\end{equation}
Where we have used that 
\begin{equation}
g_4^2 = \frac{g_5^2}{2 \p R_{x_4}} = M_\mathrm{KK} g_5^2
\end{equation}
Using the relation between the temperature and the position of the horizon \eq{eq:T} we find the potential due to flavour degrees of freedom in terms of field theory quantities is
\begin{equation}
V_\mathrm{eff} \supset \frac{\l N N_f T^4  \mathrm{Vol} \left(\mathbb{R}^3 \right) }{1728 \p M_\mathrm{KK} }(\a - \b \m_I) ^2+ \mathcal{O} \left((\a - \b \m_I) ^4 \right)
\end{equation}
We have used that $\mathrm{Vol} \left(S^4\right)= \frac{\p^2}{12}$. The potential from the flavours is quadratic, and we expect the potential due to the adjoint degrees of freedom to be periodic from an argument in \cite{Aharony:1998qu} for type IIB, generalised to type IIA in \cite{Yee:2009cd}. When the theory contains only adjoint degrees of freedom there is an infinite set of degenerate minima in the potential as a function of $\a$. This degeneracy is lifted when one includes fundamental matter and an imaginary chemical potential in the theory. The potential for the flavours only depends on the combination $\a- \b \m_I$ so if we shift $\m_I$ by $\frac{2 \p}{\b}$ the shift can be reabsorbed by $\a \to \a + 2 \p$, hence $\m_I$ is only defined up to a factor of $\frac{2 \p}{\b}$. Schematically the effective potential is
\begin{equation}
V_{\mathrm{eff}} = V_\mathrm{A} + V_\mathrm{F} \sim N^2 \left(  \min_{r \in \mathbb{Z}} \left( \a - \frac{2 \p r}{N} \right)^2 + \frac{N_f}{N} \left( \a - \b \m_I  \right)^2 \right)
\end{equation}
This is shown in figure \ref{fig:veff} for only adjoint degrees of freedom and for adjoint and fundamental degrees of freedom with zero chemical potential.
\FIGURE{
\centering
\includegraphics[width=7cm]{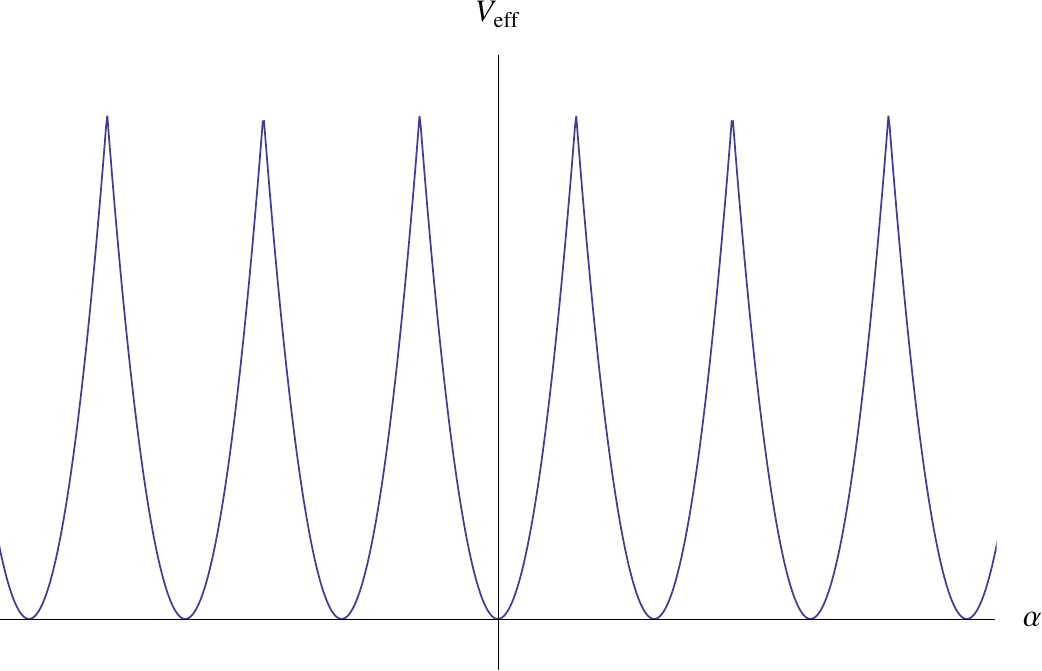}
\includegraphics[width=7cm]{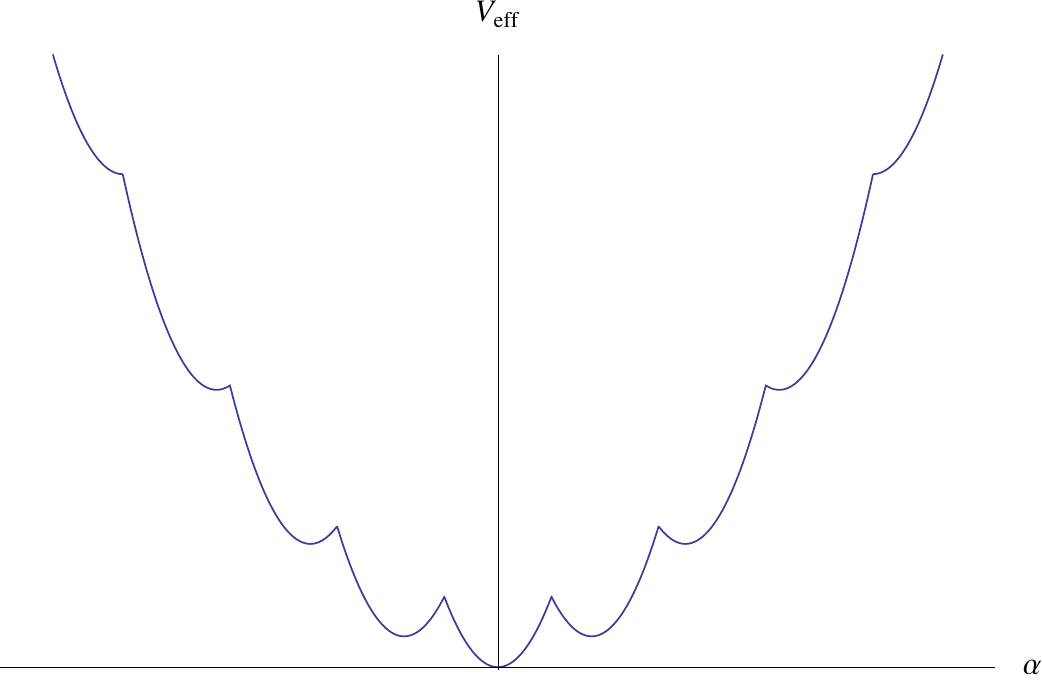}
\caption{The effective potential due to the adjoint degrees of freedom (left) and both adjoint and fundamental degrees of freedom (right) as a function of $\a$.}
\label{fig:veff}
}
For the latter it is clear there is a single global minimum and many local minima. As we increase the chemical potential a competition develops between the global minimum and an adjacent local minimum until they become degenerate, which happens when $\m_I = \m_\mathrm{RW}$ where $\m_\mathrm{RW}$ is defined in \eq{eq:murw}. Increasing the chemical potential further results in a first order transition from one minima to the next (see figure \ref{fig:veffshift}). In the new vacuum the expectation value of the phase of the Polyakov loop differs from the previous vacuum by $\frac{2 \p}{N}$.
\FIGURE{
\centering
\includegraphics[width=7cm]{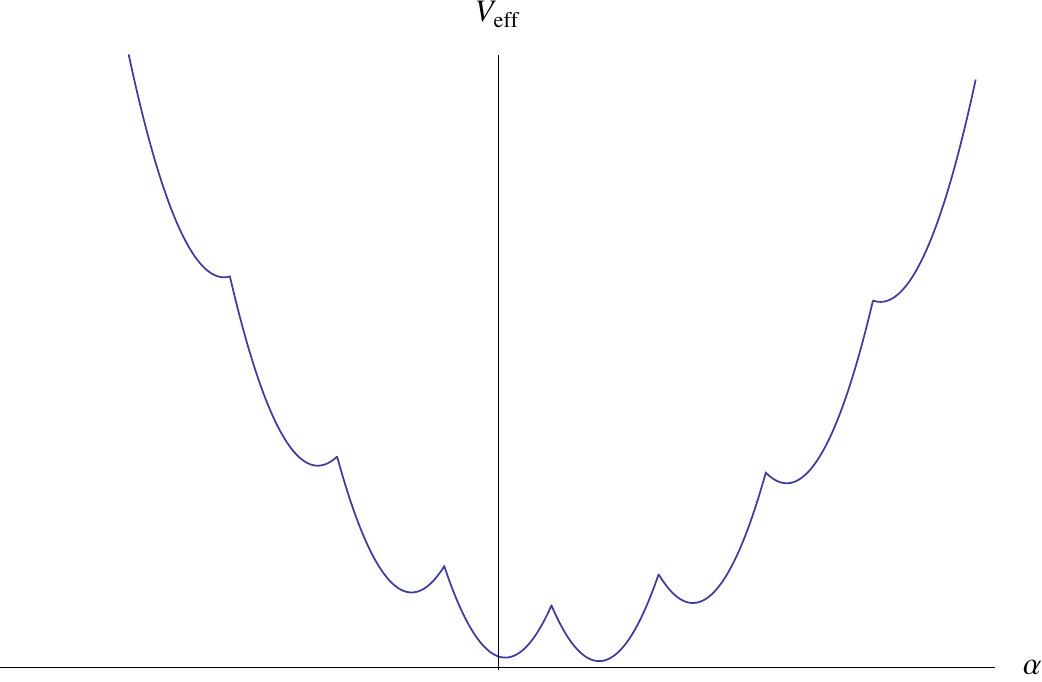}
\includegraphics[width=7cm]{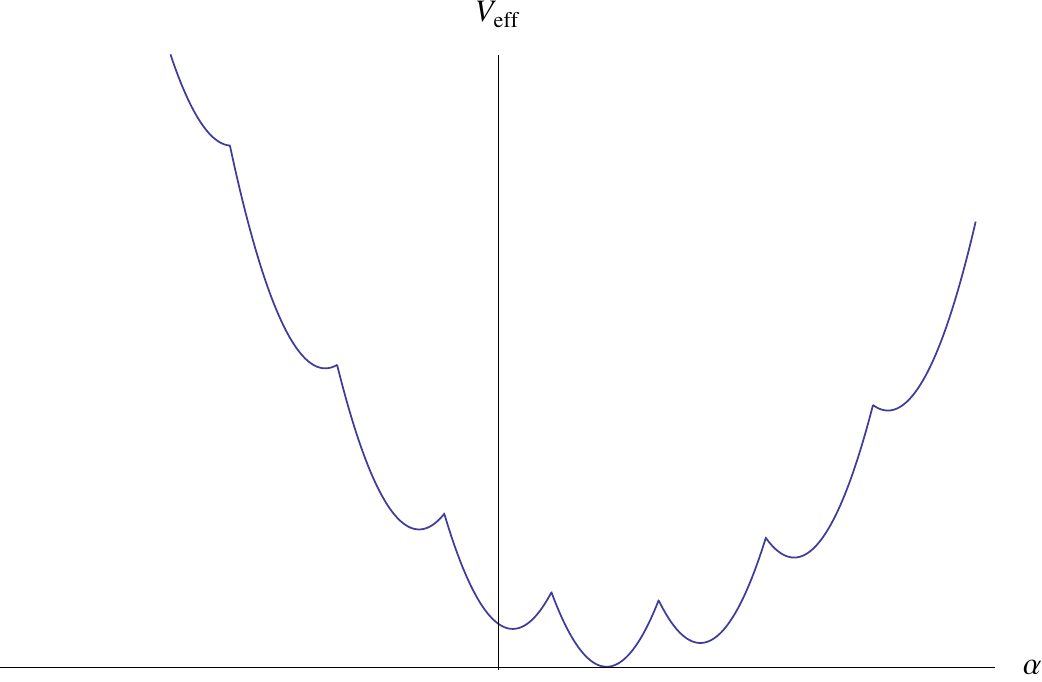}
\caption{The effective potential for adjoint and fundamental degrees of freedom when $\m_I \approx  \m_\mathrm{RW}$ (left) and $\m_I >  \m_\mathrm{RW}$ (right)}
\label{fig:veffshift}
}\\
\subsubsection{The broken chiral symmetry case}
In the language of the gravity theory this corresponds to Minkowski brane embeddings that turn around at some finite $U_0 > U_T$. In contrast to the previous case, there must be a dynamically generated quark mass in this phase, which is given by
\begin{equation}
m_Q = \frac{1}{2 \p \a'} \int_{U_T}^{U_0} dU \sqrt{|g_{tt} g_{UU}|} =  \frac{1}{2 \p \a'}\left( {U_0}-{U_T} \right) \label{eq:quarkmass}
\end{equation}
Hence, chiral symmetry is no longer a symmetry of the theory. Moreover, as shown previously $d$ must vanish for these solutions. Together with \eq{eq:bplusf} this implies that $B_{tU} + 2 \p \a' F_{tU} = 0$ and hence the potential does not depend on $\a - \b \m_I$. Since $\m_I$ is interpreted as $\lim_{U \to \infty} \left[A\left(U\right)\right]$ we are still able to introduce a chemical potential via the introduction of a constant gauge field, but the physics remains smooth for all values of this chemical potential. 

This analysis is also applicable to the low temperature geometry where this type of D8 - $\overline{\mathrm{D8}}$ embedding is the only one allowed. The conclusion is the same and the potential does not depend on $\a - \b \m_I$. 
\subsection{The Phase Diagram}
At low temperatures the only solutions allowed are Minkowski embeddings so the phase diagram has no phase transitions. Above a certain temperature there is a transition from \eq{eq:coldmetric} to \eq{eq:metric} and both types of solution are then allowed. We will compute the phase diagram of the theory at high temperature by considering the difference in the DBI actions of the two types of solution, dual to the grand canonical ensemble in the gauge theory. After applying the conditions for each of the solutions derived above and substituting in using the equations of motion we find
\begin{align}
S_\mathrm{Mink} &= \frac{\sqrt{R^3} N_f T_{\mathrm{D}8} \mathrm{Vol} \left(S^4\right)  \mathrm{Vol} \left(\mathbb{R}^3\right) \b}{g_s} \int_{U_0}^\infty dU U^5 \sqrt{\frac{U^3 f(U)}{U^8 f(U) -U_0^8 f(U_0)}}\\
S_\mathrm{BH} &= \frac{\sqrt{R^3} N_f T_{\mathrm{D}8} \mathrm{Vol} \left(S^4\right) \mathrm{Vol} \left(\mathbb{R}^3 \right)\b}{g_s} \int_{U_T}^\infty dU \frac{U^5}{\sqrt{U^5 - \frac{\tilde{d}^2}{R^3}}}
\end{align}
hence we define
\begin{align}
\D S &= \frac{1}{\mathcal{C}}\left(S_\mathrm{Mink} - S_\mathrm{BH} \right) \nonumber\\
&= \int_{U_0}^\infty dU U^5 \left( \sqrt{\frac{U^3 f(U)}{U^8 f(U) -U_0^8 f(U_0)}} - \frac{1}{\sqrt{U^5 - \frac{\tilde{d}^2}{R^3}}}\right) - \int_{U_T}^{U_0} dU \frac{U^5}{\sqrt{U^5 - \frac{\tilde{d}^2}{R^3}}} \label{eq:deltas} \\
\mathcal{C} &=  \frac{\sqrt{R^3} N_f T_{\mathrm{D}8} \mathrm{Vol} \left(S^4\right)  \mathrm{Vol} \left(\mathbb{R}^3\right) \b}{g_s}
\end{align}
When $\D S = 0$ there is a phase transition. It is worth making some remarks on the method used in \cite{Horigome:2006xu} to compute the phase diagram of the Sakai-Sugimoto model in the presence of a real chemical potential. The following dimensionless coordinates were used to eliminate $U_0$ from $\D S$
\begin{equation}
u_T = \frac{U_T}{U_0} \qquad u = \frac{U}{U_0}  \qquad c^2 = \frac{\tilde{d}^2}{R^3 U_0^5}
\end{equation}
where Wick rotating $c$ would recover the real chemical potential result. In terms of these coordinates we have
\begin{align}
\D S_\mathrm{Dimensionless} = \int_{1}^\infty du  u^5 \left( \sqrt{\frac{u^3 f(u)}{u^8 f(u) -f(1)}} - \frac{1}{\sqrt{u^5 - c^2}} \right) - \int_{u_T}^{1} du \frac{u^5}{\sqrt{u^5 - c^2}} \label{eq:deltas_dl}
\end{align}
Henceforth we shall refer to the difference in the actions in dimensionless coordinates of \cite{Horigome:2006xu} as $\D S_\mathrm{Dimensionless}$, while in the dimensionful coordinates we shall denote is $\D S$. Evaluating \eq{eq:deltas_dl} for $c=0$ or for zero chemical potential we find figure \ref{fig:zeromudeltas}.
\FIGURE{
\centering
\includegraphics{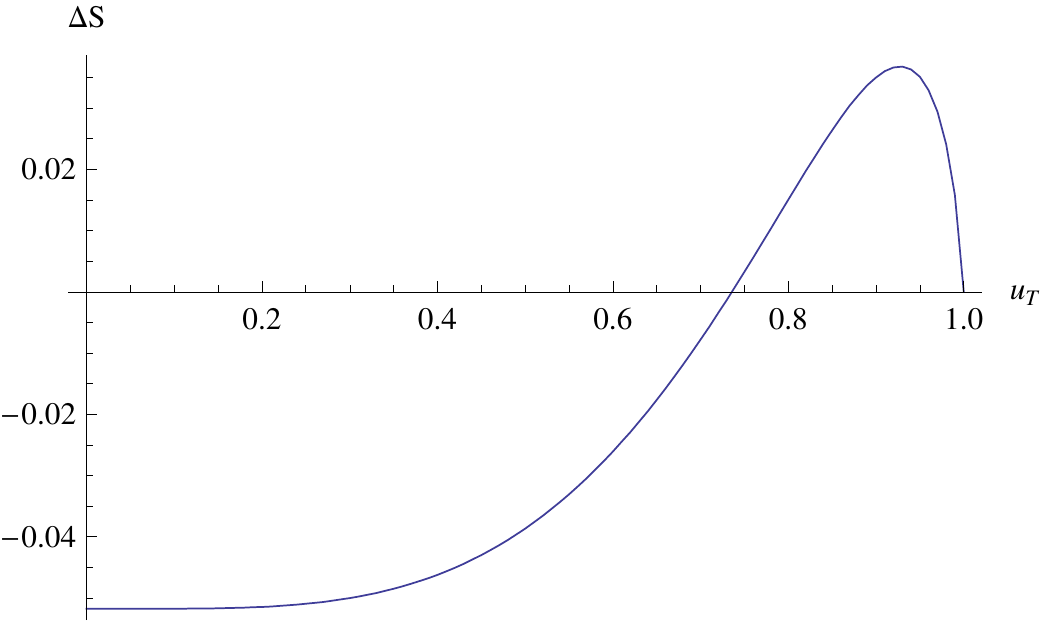} 
\caption{The plot of $\D S$ against $u_T= \frac{U_T}{U_0}$ for zero chemical potential}
\label{fig:zeromudeltas}
}
There are two zeros, one of which occurs at approximately $u_T = 0.74$ which corresponds to the $\m = 0$ limit of the real chemical potential phase transition line. This was originally computed in \cite{Aharony:2006da,Parnachev:2006dn}. The second is at $u_T = 1$ which, when one considers a real chemical potential is lifted and does not contribute to the phase diagram. When we consider an imaginary chemical potential however, this zero persists and would lead to a second phase transition line on the phase diagram. We will show that the phase transition line that this zero leads to should not, in fact be present on the phase diagram. Consider the expression relating the dimensionless horizon position and the temperature (from equation \eq{eq:T})
\begin{equation}
T = \frac{3}{4 \p} \sqrt{\frac{U_0 u_T}{R^3}} \label{eq:T_dimensionless}
\end{equation}
The asymptotic separation of the D8 - $\overline{\mathrm{D}8}$ pair is $L$ and is given by; 
\begin{align}
L &= 2 \lim_{U \to \infty} x_4(U) \nonumber \\
 &= 2 \int_{U_0}^\infty dU x_4' (U) \nonumber\\
 &= 2 \sqrt{R^3} \int_{U_0}^\infty dU \frac{1}{\sqrt{U^3 f(U) \left(\frac{U^8}{U_0^8} \frac{f(U)}{f(U_0)} -1\right)}} \nonumber \\
& = \sqrt{\frac{R^3}{U_0}} F\left(u_T\right) \label{eq:L}
\end{align}
where
\begin{equation}
F\left(u_T\right)= 2 \int_1^\infty du \frac{1}{\sqrt{u^3 f(u) \left(u^8 \frac{f(u)}{f(1)} -1\right)}}
\end{equation}
Note that for black hole embeddings $L$ is arbitrary, so we consider Minkowski embeddings by plugging in for $x_4'$ using \eq{eq:eom}. Using \eq{eq:L} to eliminate $U_0$ from \eq{eq:T_dimensionless}
\begin{equation}
T L = 2 \sqrt{u_T} F\left(u_T\right) \label{eq:Tbar}
\end{equation}
$TL$ as a function of $u_T$ is not a monotonically increasing function and has a global maximum, it is shown in figure \ref{fig:TL}. 
\FIGURE{
\centering
\includegraphics[width=10cm]{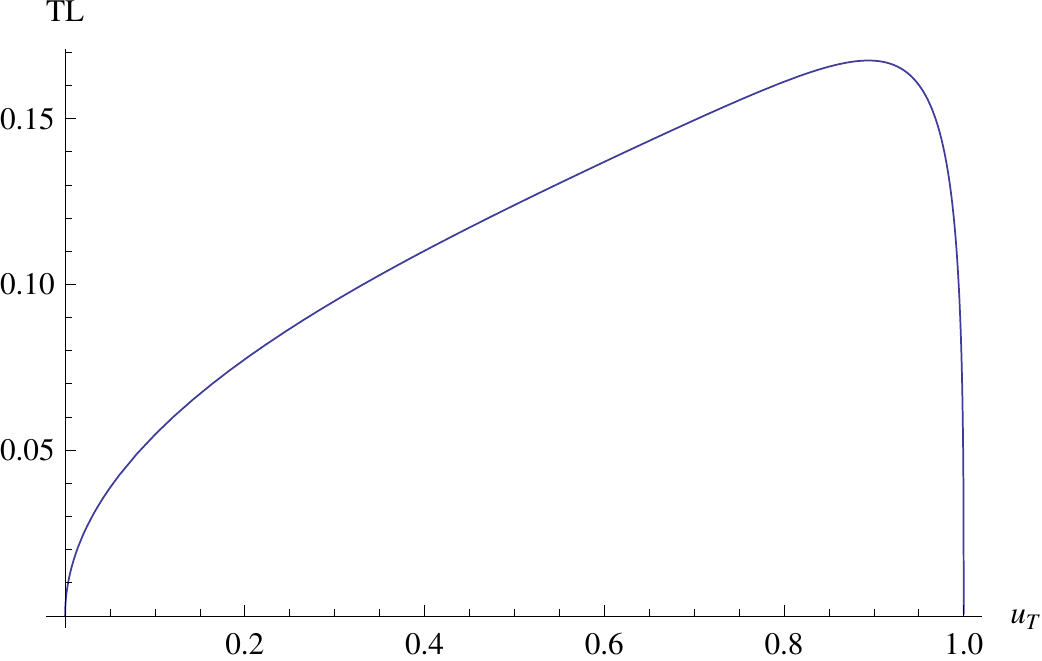}
\caption{$TL$ as a function of $u_T$}
\label{fig:TL}
}
Suppose we consider a fixed $T$ and $L$. There are two allowed Minkowski embeddings that have the same value of $TL$ but have different values of $u_T$. For a fixed $u_T$ however there is only one Minkowski embedding with a particular allowed value of $TL$. We wish to compute the phase diagram by considering the action for all allowed solutions at each $TL$ and chemical potential (there are three - one black hole embedding and the two Minkowski embeddings), the one that has the lowest action dominates. Using $u_T$ as a parameter to do this is problematic because for each $u_T$ we are only considering one of the Minkowski embeddings allowed at that particular $TL$. Computing the phase diagram using a parameter that only considers a subset of allowed solutions leads to a phase diagram with a phase transition line that would not appear were we to consider the full set of allowed solutions.

The analysis of $\D S_\mathrm{Dimensionless}$ has one important consequence however: we note that $\D S_\mathrm{Dimensionless}$ and the combination $TL$ are functions of $u_T$ only (at zero chemical potential), therefore the action is a function of $TL$, rather than $T$ and $L$ individually. Consider now the chemical potential. Rewriting \eq{eq:mu} using the dimensionless coordinates we have
\begin{equation}
\a - \b \m_I = \b U_0 \int_{u_T}^\infty du \frac{c}{\sqrt{u^5 -c^2}} 
\end{equation}
eliminating $U_0$ as before we have 
\begin{equation}
\frac{L^2 }{\b R^3} \left(\a - \b \m_I\right) =  F\left(u_T\right) \int_{u_T}^\infty du \frac{c}{\sqrt{u^5 - c^2}} \label{eq:mubar}
\end{equation}
hence, the combination of the chemical potential and the expectation value of the phase of the Polyakov loop is a function of $u_T$ and $c$, and vanishes when $c \to 0$. There are two scales on the gravity side ($c$ and $u_T)$ that correspond to two meaningful scales on the gauge thoery side; $TL$ and $\frac{L^2 }{\b R^3} \left(\a - \b \m_I\right)$, which we will denote by $\overline{T}$ and $\overline{\m}$ respectively. We will plot the phase diagram of the model as a function of these two parameters. The phase diagram is computed by finding the zeros of $\D S$ \eq{eq:deltas} as a function of $U_T$, $\tilde{d}$ for a fixed $U_0$ and converting these coordinates to $\overline{T}$ and $\overline{\m}$. We note that repeating this computation for various values $U_0$ reproduces the same phase diagram, which is shown in \ref{fig:phase_ss}.
\FIGURE{
\centering
\includegraphics[width=10cm]{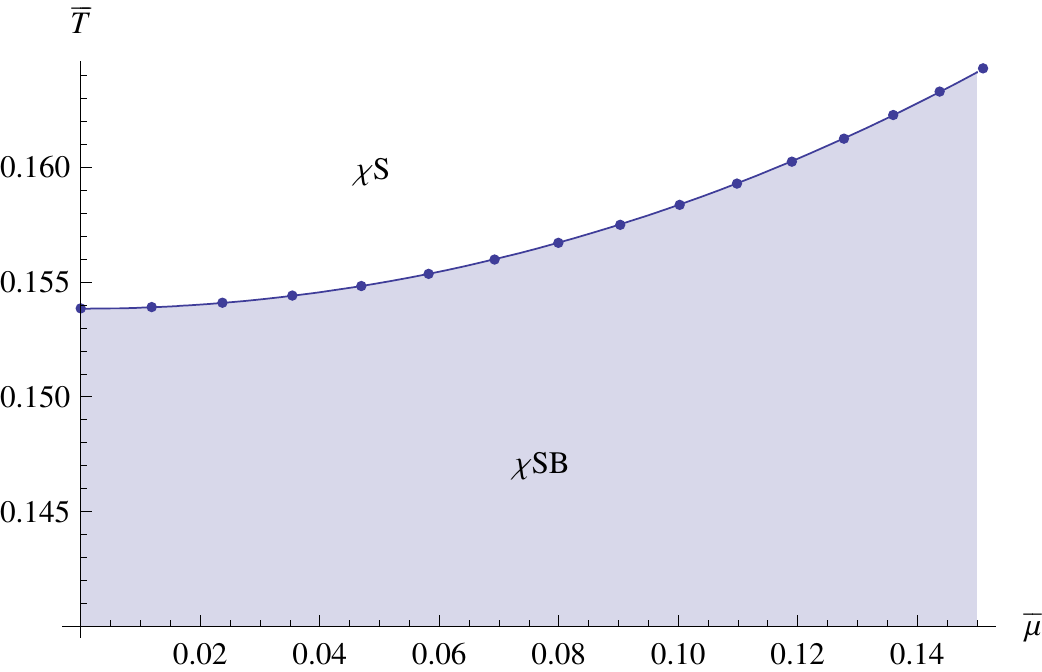}
\caption{The phase diagram of the Sakai-Sugimoto model with an imaginary chemical potential as a function of the dimensionless temperature and chemical potential.}
\label{fig:phase_ss}
}
The blue line is the first order phase transition line that extends from the phase transition point found in \cite{Aharony:2006da,Parnachev:2006dn} for zero chemical potential. The shaded area of the plot is the area where the chiral symmetry broken phase dominates and the unshaded area is the one where chiral symmetry is restored. 

The issue we have noted regarding the use of the dimensionless coordinates to study the phase structure of the model was not probelematic for real chemical potential. We will now show why this was the case. Consider $\D S_\mathrm{Dimensionless}$ for the theory when $\overline{\m}$ is small. Expanding in powers of $c$ and similarly expanding equation \eq{eq:mubar} we may capture the leading behaviour of $\D S_\mathrm{Dimensionless}$ as a function of $\overline{\m}$
\begin{align}
\D S_\mathrm{Dimensionless} &\approx \D S_\mathrm{Dimensionless} \big|_{c = 0} -\frac{9 \overline{\mu} ^2}{52 F\left(u_T\right)^2 \sqrt{u_T^{7}}} + \mathcal{O}\left(\overline{\mu} ^4\right) \nonumber\\
&\approx \D S_\mathrm{Dimensionless} \big|_{c = 0} -\frac{9 \overline{\mu} ^2}{52 \overline{T}^2 \sqrt{u_T^{5}}} + \mathcal{O}\left(\overline{\mu} ^4\right)
\end{align}
where we have used \eq{eq:Tbar} to exchange $F\left(u_T\right)$ for $\overline{T}$. $\D S_\mathrm{Dimensionless} \big|_{c = 0}$ must be evaluated numerically and is shown in figure \ref{fig:zeromudeltas}. The factors of $u_T$ in the denominator of the above expression are problematic at first sight, however, we note that the limited range of $u_T$ from \eq{eq:range} implies $u_T^5 \geq c^2$ so for a fixed value of $c > 0$, $u_T$ must be strictly greater than zero and the first correction term never diverges. When $u_T$ is close to 1, $\D S_\mathrm{Dimensionless}\big|_{c=0}$ is small and positive. For an imaginary chemical potential the first correction term that appears at order $\overline{\m}^2$ is negative leading to the second zero of $\D S_\mathrm{Dimensionless}$. Were we to Wick rotate $\overline{\m}$ so that we are considering the theory with a real chemical potential the sign of this correction term would be positive, so $\D S_\mathrm{Dimensionless}$ does not vanish when $u_T$ is close to 1 for a finite but small real $\overline{\m}$.

Performing a similar expansion for small $\tilde{d}$ on $\D S$ \eq{eq:deltas} is also instructive. Using an expansion of \eq{eq:mu} to eliminate $\tilde{d}$ and \eq{eq:T} to eliminate $U_T$ we find
\begin{equation}
\D S \approx \D S\big|_{\tilde{d} = 0} -\frac{64}{9} \left(\a - \b \m_I \right)^2  \pi ^5 R^{\frac{9}{2}} T^5 \alpha'^2 + \mathcal{O} \left( \left(\a - \b \m_I \right)^4 \right) 
\end{equation}
We could in principle restore the dimensionful coefficient and rewrite $\D S$ using field theory quantities, however it is not very enlightening to do so. It is important to note that the first correction term does not depend on $U_0$ and the equations relating $T$ to $U_T$ and $\left(\a - \b \m_I \right)$ to $\tilde{d}$ contain no dependence on $U_0$ so the above expression for $\D S$ contains no hidden factors of $U_0$. Therefore the system must be well behaved when $U_0$ approaches $U_T$, which when one considers the model using the dimensionless coordinates discussed above corresponds to the $u_T \to 1$ regime. This analysis also shows the action depends on the square of the chemical potential, ensuring the analyticity of the phase diagram as a function of $\left(\a - \b \m_I \right)^2$.

In summary, when comparing the actions of the two different types of embedding it is important to work in a coordinate system where there is a one to one correspondence between the position of a horizon and a temperature. 

In the low temperature phase when $\overline{T} < \overline{T}_d$ where $\overline{T}_d$ is the deconfinement temperature chiral symmetry is broken and the glue in the theory is confined. Above the deconfinement temperature in the chiral symmetry broken phase the glue is deconfined and the theory contains mesons. In the chiral symmetry restored phase the mesons melt and the fundamental degrees of freedom in the theory are quarks and gluons. The quarks are also massless due to the chiral symmetry restoration. Including the Roberge-Weiss lines the phase diagram of the theory is given in figure \ref{fig:phase_final}. The equation of the phase transition line is
\begin{equation}
\overline{T} = 0.154+0.447 \overline{\m}^2 + \mathcal{O}\left(\overline{\m}^4\right)
\end{equation}
The equation for $\overline{T}$ behaves as expected, ie, is a function of even powers of $\overline{\m}$. 
\FIGURE{
\centering
\includegraphics[width=12cm]{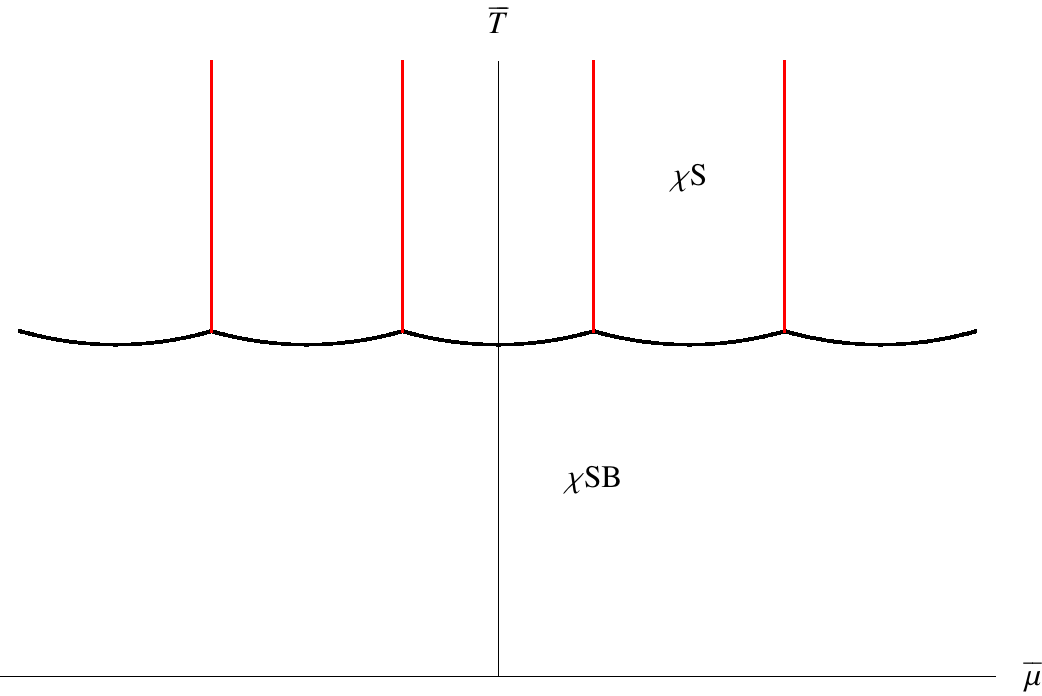}
\caption{The phase diagram of the Sakai-Sugimoto model with an imaginary chemical potential. The black line extends from the line found for the real chemical potential case. The red lines are the Roberge-Weiss lines which appear in the chiral symmetry restored phase. Since the equation of the phase transition line is a function of $\overline{\m}^2$ the second derivative of the phase transition line is strictly positive so even though the Roberge-Weiss lines appear at order $\frac{1}{N}$ the chiral symmetry breaking transition does show curvature between the Roberge-Weiss lines.}
\label{fig:phase_final}
}
\subsection{Pressure due to Flavours}
We will compute the pressure holographically by studying the DBI action as a function of temperature. From basic thermodynamics
\begin{equation}
F = T dS - P dV + \m dN
\end{equation}
The pressure is given by
\begin{equation}
P = - \pd{F}{V}
\end{equation}
holographically, the DBI action is dual to the free energy density in the grand canonical ensemble, therefore it is also related to the pressure. The DBI action is
\begin{equation}
S_\mathrm{DBI} = \int dU d\t d^3 x d\W_4 \mathcal{L}_\mathrm{DBI}
\end{equation}
The integrals over the field theory directions contribute a volume of $\mathbb{R}^3$ and a factor of $\b$, because $\mathcal{L}_\mathrm{DBI}$ is not a function of these directions. One may go from the energy density to the pressure by dividing through by these factors or we may directly compute the pressure by evaluating 
\begin{equation}
P = -\int dU d\W_4 \mathcal{L}_\mathrm{DBI}
\end{equation}
\subsubsection{Black Hole Embeddings - Zero Density}
When the density is zero the pressure takes a simple form
\begin{equation}
P_\mathrm{BH} =-\mathcal{N} \int_{U_T}^\infty dU U^{\frac{5}{2}}
\end{equation}
we have relabelled the overall constant in front of the pressure as 
\begin{equation}
\mathcal{N} = \frac{\p^2 \sqrt{R^3} N_f T_{\mathrm{D}8}}{12 g_s}
\end{equation}
Regulating the integral and adding a counterterm to cancel the divergence as $\L \to \infty$ we have
\begin{equation}
P_\mathrm{BH} = - \lim_{\L \to \infty}\mathcal{N}  \left(\frac{2 U^{\frac{7}{2}}}{7} \bigg|_{U_T}^\L - \frac{2 \L^{\frac{7}{2}}}{7} \right)
\end{equation}
Hence
\begin{equation}
P_\mathrm{BH} = \frac{2 }{7 } \mathcal{N}  U_T^{\frac{7}{2}}
\end{equation}
Using the expression for the temperature given in \eq{eq:T} we can show
\begin{equation}
P_\mathrm{BH} =\frac{2^4}{7} \frac{\left(2 \p\right)^9 T_{\mathrm{D}8} N_f}{3^8 g_s}  R^{12} T^7
\end{equation}
Using the expressions given in section \ref{sec:pressure} we find the pressure in terms of field theory quantities is
\begin{equation}
P_\mathrm{BH} =\frac{\l^3}{7.3^8} \frac{N_f N}{2 \p}  T^4  \left( \frac{T}{M_\mathrm{KK}}\right)^3
\end{equation}
where $\l = g_4^2 N$. This expression is quite surprising, in particular the pressure is proportional $T^7$.
\subsubsection{Non Zero Density}
Expanding $P_\mathrm{BH}$ for small $\tilde{d}$ we have
\begin{equation}
P_\mathrm{BH} =\mathcal{N}  \int_{U_T}^\infty dU \left(U^{5/2}+\frac{\tilde{d}^2}{2 R^3 U^{5/2}} + \mathcal{O}\left(\tilde{d}^4\right) \right)
\end{equation}
Evaluating the integral as before, showing only the first order correction for non zero $d$
\begin{equation}
P_\mathrm{BH} \supset -\mathcal{N} \left( \frac{3^{2} \tilde{d}^2 }{(4 \p)^3 R^{\frac{15}{2}} T^3} + \mathcal{O}\left(\tilde{d}^4\right) \right)
\end{equation}
For small density we can expand the expression for the chemical potential (assuming the phase of the Polyakov loop vanishes)
\begin{align}
\m_I &= -\frac{1}{2 \p \a'} \int_{U_T}^\infty dU \frac{\tilde{d}}{\sqrt{U^5 R^3 -\tilde{d}^2}}\\
\Rightarrow d &\approx \frac{8 \pi^5 \left(2 \p \a' \right)^2 N_f  T_\text{D8}  }{3^3 g_s}R^6 T^3 \mu_I  + \mathcal{O} \left(\m_I^3 \right)
\end{align}
where we have restored the correct dimensionality of the density. In terms of field theory quantities, the density in terms of the chemical potential is 
\begin{equation}
d \approx \frac{N_f N}{2 \pi} \frac{ T^3 \lambda  \mu_I }{2^4 3^3 M_\mathrm{KK}} + \mathcal{O} \left(\m_I^3 \right)\label{eq:smallmu}
\end{equation}
The first order correction depends on the square of the chemical potential and it is
\begin{equation}
-\frac{ N_f N}{2 \p} \frac{\l }{2^5 3^3 } \frac{T^3 \mu ^2}{M_\mathrm{KK}}+\mathcal{O}\left(\m^4 \right)
\end{equation}
Together, the leading order behaviour of the pressure is
\begin{equation}
P =  \frac{1}{3^3} \frac{N_f N}{2 \p}  T^4   \left(\frac{\l^3}{7.3^5} \left( \frac{T}{M_\mathrm{KK}}\right)^3 - \frac{\l }{2^5 } \frac{\mu ^2}{T M_\mathrm{KK}}  + \mathcal{O}\left(\m^4 \right)\right)
\end{equation}
Since the pressure is a mass dimension 4 object, the expression for the pressure will always be proportional to $T^4$ multiplied by a dimensionless function of the scales in the problem, in this case $T, \m_I$ and $M_\mathrm{KK}$. An expected feature of the expression for the pressure is that the leading order term dominates when the temperature is very large but, as commented previously, the pressure computed here does not correlate with what is known in QCD. The action for the black hole embeddings can be evaluated exactly in terms of a hypergeometric function, and it is
\begin{align}
P =\frac{N_f N \lambda ^3}{7. 2^2 3^8  \pi ^{\frac{5}{2}} (\mathcal{T}-1) \mathcal{T}^{\frac{7}{10}}} \left( \frac{T}{ M_\mathrm{KK}}\right)^3 T^4 \Bigg( &(-1)^{\frac{3}{10}} (\mathcal{T}-1) \G\left(\frac{3}{10}\right) \Gamma\left(\frac{6}{5}\right) -\sqrt{\pi } \sqrt{\mathcal{T}-1} \mathcal{T}^{\frac{6}{5}} + \nonumber\\
&+\sqrt{\pi } (\mathcal{T}-1)^{\frac{3}{2}} \mathcal{T}^{\frac{1}{5}} {}_2F_1\left(\frac{7}{10},1,\frac{6}{5},\mathcal{T}\right)\Bigg)
\end{align}
where
\begin{equation}
\mathcal{T} =\frac{N_f^2 N^2}{\left(2 \p\right)^2} \frac{  \l^4}{ 2^4 3^{12}} \frac{T^{10}}{M_\mathrm{KK}^4 d^2}
\end{equation}
Since this formalism captures the strong coupling behaviour of the field theory expanding for large 't Hooft coupling is a natural thing to do
\begin{equation}
P \approx T^4 \left( \frac{N_f N}{2 \pi} \frac{  \lambda^3}{7. 3^8 }\frac{T^3}{M_\mathrm{KK}^3} - \frac{2 \pi}{N_f N }\frac{6^3 }{7  \lambda }\frac{d^2 M_\mathrm{KK}}{T^7} + \mathcal{O} \left(\frac{1}{\l^5} \right)\right)
\end{equation}
When one substitutes in for small $d$ in terms of $\m_I$ using \eq{eq:smallmu} the same result as for the small density expansion is obtained.
\subsubsection{Minkowski Embeddings}
The integral for the action for Minkowski embeddings cannot be evaluated in closed form. The pressure for the Minkowski embeddings is
\begin{equation}
P_\mathrm{Mink} =-\mathcal{N}  \int_{U_0}^\infty dU U^5 \sqrt{\frac{U^3 f(U)}{U^8 f(U) -U_0^8 f(U_0)}}
\end{equation}
Rewriting:
\begin{equation}
P_\mathrm{Mink} =-\mathcal{N} \int_{U_0}^\infty dU \frac{U^{\frac{5}{2}}}{ \sqrt{1 - \frac{U_0 f\left(U_0\right)}{U f\left(U \right)} } }
\end{equation}
Replacing $U_0$ using
\begin{equation}
2 \p \a' m_q = M = U_0 - U_T
\end{equation}
and expanding in powers of $M$ (note, since the lower limit of integration is a function of $M$ we expand the integrand, evaluate the integral and then expand the result. The first order correction only receives contributions from the first order of the expansion of the integrand, while the corrections at order $M^2$  contain contributions from higher orders of the expansion of the integrand.)

In the zero mass limit the pressure for Minkowski embeddings captures the same behaviour of the chiral symmetry restored phase so we choose to only consider the terms proportional to $M$ as the leading behaviour. To $\mathcal{O}\left(M^2\right)$ we have
\begin{equation}
P_\mathrm{Mink} = -\mathcal{N}\int_{U_0}^\infty dU \left(\frac{3 U_T^7 M}{2 U^{5/2} \left(U^3-U_T^3\right)}+\frac{9 U_T^6 \left(8 U^8-8 U^5 U_T^3+3 U_T^8\right) M^2}{8 U^{15/2} \left(U^3-U_T^3\right)^2} + \mathcal{O}\left(M^3\right) \right)
\end{equation}
Once we have performed the integral the leading behaviour of the pressure is
\begin{equation}
P_\mathrm{Mink} = -\frac{\mathcal{N}}{20} M  U_T^{5/2} \left( 5 \log \left(3. 2^4  \left(\frac{U_T}{M}\right)^2\right)+5 \sqrt{3} \pi  -12 \right)
\end{equation}
using the above expressions to rewrite in terms of field theory quantities we have
\begin{equation}
P_\mathrm{Mink}= -\frac{\l^2}{5.2^4 3^6} \frac{ N_f N}{ 2 \p} T^4 \frac{m_q T}{ M_\mathrm{KK}^2} \left(5 \log\left(\frac{64 \l^2}{27} \frac{ T^4 }{ m_q^2 M_\mathrm{KK}^2}\right)+5 \sqrt{3} \pi-12\right)
\end{equation}
therefore the leading behaviour is proportional to $T^5 \frac{m_q}{ M_\mathrm{KK}^2}\log \left( \frac{ T^4 }{ m_q^2 M_\mathrm{KK}^2} \right)$.
\subsubsection{Hadron Resonance Gas Model}
Since the results we have here are for a system which contains bound states of quarks and strong coupling it is natural to compare to the hadron resonance gas model of QCD \cite{Karsch:2003vd}. This is a model of non interacting hadronic and mesonic resonances, which is a good model of QCD at low temperatures since the quarks are confined and are no longer the physical degrees of freedom of the theory. The pressure contribution of a single particle in this model is given by
\begin{equation}
\b^4 P_i = \frac{g_i}{2 \p^2}\sum_{k=1}^\infty \left(-\eta\right)^{k+1}\frac{\left(\b m_i\right)^2}{k^2} K_2 \left(k \b m_i \right)
\end{equation}
$\b$ is the inverse temperature, $g_i$ is a degeneracy factor and $\eta = 1$ for bosons and $-1$ for fermions. $K_2$ is a modified Bessel function of the second kind. Assuming the mass is small we have 
\begin{equation}
\b^4 P_i = \frac{g_i}{2 \p^2 }\sum_{k=1}^\infty\left( \frac{2 (-\eta )^{1+k}}{k^4}+\frac{ \beta ^2 (-\eta )^k \eta  m_i^2}{2 k^2 }+\mathcal{O} \left(m^4 \right) \right)
\end{equation}
evaluating the sum
\begin{equation}
\b^4 P_i = \frac{\eta g_i }{2 \pi ^2} \left(-2 \text{Li}_4\left(-\eta \right)+\frac{m_i^2 \beta ^2}{2} \text{Li}_2\left(-\eta \right)+\mathcal{O} \left(m_i^4 \right)  \right)
\end{equation}
Note that $\text{Li}$ is a polylogarithm function that contributes a numerical factor when we plug in for $\eta$. Note that the first order correction to the pressure for non zero quark mass is 
\begin{equation}
P \sim T^4 \frac{m^2}{T^2}
\end{equation}
While we note that the Sakai-Sugimoto is very different from QCD, being 5 dimensional and having a large number of colours etc., the pressure from the hadron resonance gas model exhibits very different behaviour to that found in the Sakai-Sugimoto model above. 
\section{Defect Theories}
\label{sec:defect}
The work of \cite{Aarts:2010ky} showed that when one considers $\mathcal{N}=4$ super Yang-Mills theory with quenched $\mathcal{N}=2$ flavours at strong coupling, dual to type IIB supergravity on AdS$_5 \times$ S$^5$ with probe D7 branes, in the presence of an imaginary chemical potential the resulting phase diagram contains only first order phase transition lines. Therefore, the points where the Roberge-Weiss lines meet the deconfinement transition are triple points. It would be interesting to find a model where the deconfinement line is not of the first order and one way that could possibly be the case is in a defect theory. This type of background was discussed in a very general way in \cite{Benincasa:2009be}, but we limit ourselves to the background that comes from the backreaction of D3 branes in type IIB supergravity. To include flavours we put $N_f$ D$(7-2k)$ branes into the AdS$_5 \times$ S$^5$ background. $k$ is the codimension of the defect which we take as $k=1$ ($k=2$) representing a bulk theory containing $N_f$ D5 (D3-$\overline{\mathrm{D}3}$) probe branes in the geometry that results from $N$ D3 branes where $N \to \infty$. This bulk theory corresponds to a field theory where the fundamental matter lives on a $2+1$ ($1+1$) dimensional submanifold of the $3+1$ dimensional boundary. The adjoint fields live in the full $3+1$ dimensions of the boundary. $k=0$ would correspond to having no defect, ie the D3 /D7 system. In order for the theory to be retain $\mathcal{N}=2$ supersymmetry, the brane must cover $3-k$ spatial dimensions of the boundary and $3-k$ directions of the S$^5$ in addition to the time direction and the radial direction. This type of theory has been discussed extensively in the literature. See for example \cite{DeWolfe:2001pq, *Myers:2008me,*Wapler:2009tr,*Hung:2009qk}.

The metric corresponding to $\mathcal{N}=4$ super Yang-Mills at high temperature having Wick rotated to Euclidean signature is
\begin{equation}
ds^2 = \left(\frac{r}{R}\right)^2 \left(h(r) d\t^2 + \d_{ij} dx^i dx^j \right)+ \left(\frac{R}{r}\right)^2 \left(\frac{dr^2}{h(r)}+r^2 d\W^2_5 \right)
\end{equation}
with 
\begin{equation}
h(r) = 1- \left(\frac{r_h}{r}\right)^4
\end{equation}
Rescaling the radial coordinate using
\begin{equation}
\frac{d\s}{\s} = \frac{dr}{r \sqrt{h(r)}}
\end{equation}
leads to the metric
\begin{equation}
ds^2 = \left(\frac{\s}{R}\right)^2 h_+(\s) \left(\left(\frac{h_-(\s)}{h_+(\s)}\right)^2 d\t^2 + \d_{ij} dx^i dx^j \right)+ \left(\frac{R}{\s}\right)^2 \left(d\s^2+\s^2 d\W^2_5 \right)
\end{equation}
where $i$ and $j$ run from 1 to 3 and
\begin{equation}
h_\pm (\s) = 1 \pm \left(\frac{\s_h}{\s}\right)^4
\end{equation}
For convenience we choose to rewrite the metric of the transverse coordinates as 
\begin{equation}
d\s^2+\s^2 d\W^2_5 = d\r^2 + dy^2 + \r^2 d\W_{3-k}^2 + y^2 d \W^2_{1+k}
\end{equation}
The Hawking temperature in this coordinate system is given by
\begin{equation}
T=\frac{\sqrt{2} }{\p R} \frac{\s_h}{R} \label{eq:defectT}
\end{equation}
Assuming that the position of the brane in the $y$ direction is a function of the radial coordinate only, the induced metric on the brane is given by
\begin{align}
ds_{\mathrm{D}(7-2k)}^2= \left(\frac{\s}{R}\right)^2 h_+(\s) \left( \left(\frac{h_-(\s)}{h_+(\s)}\right)^2  d\t^2 + \d_{\tilde{i} \tilde{j}} dx^{\tilde{i}} dx^{\tilde{j}} \right) + \nonumber\\ + \left(\frac{R}{\s}\right)^2 \left( \left(1+ y'(\r)^2 \right) d\r^2 + \r^2 d\W_{3-k}^2 \right)\label{eq:defectmetric}
\end{align}
where $\tilde{i}$ and $\tilde{j}$ run from 1 to $3-k$ and $\s^2 = \r^2 + y^2$. One may now write the DBI action for the D$(7-2k)$ brane and solve the resulting equation of motion to get the profile of the brane, from which we can compute the free energy in the field theory. We note that in the case of field theories dual to D3 brane geometries the resulting boundary theory is always conformal so the equation of motion for the dilaton is solved by a constant for all of the defect theories we consider.

\subsection{Brane Embedding}
From the induced metric of a D$(7-2k)$ brane \eq{eq:defectmetric} we can write the DBI action of $N_f$ branes including a worldvolume gauge field living on them in the same way as for the Sakai-Sugimoto model above
\begin{equation}
S_{\mathrm{D}(7-2k)} = T_k N_f \b V_k \mathrm{Vol}(\mathrm{Defect}) \int d\r \r^{3-k} h_+^{\frac{3-k}{2}} \sqrt{\frac{h^2_-}{h_+}\left(1+y'^2\right)+\left(B +2 \p \a' F \right)^2} \label{eq:defectaction}
\end{equation}
Note that $T_k$ is the tension of the branes and $\b$ is the inverse temperature. $V_k$ is the volume of an $S^{3-k}$. Since the action only depends on the derivative of the gauge potential and not the gauge potential itself there is an integral of motion
\begin{equation}
\pd{\mathcal{L}}{F} = c_F \Rightarrow \frac{c_F}{2 \p \a'  T_k N_f V_k} = \tilde{c}_F =  \frac{\left(B +2 \p \a' F \right) \r^{3-k} h_+^{\frac{3-k}{2}}}{\sqrt{\frac{h^2_-}{h_+}\left(1+y'^2\right)+\left(B +2 \p \a' F \right)^2}} \label{eq:cfdef}
\end{equation}
hence:
\begin{equation}
\left(B +2 \p \a' F \right)^2 =\frac{h^2_-}{h_+} \frac{\tilde{c}_F^2 \left(1+y'^2\right)}{\r^{2(3-k)} h_+^{3-k} -\tilde{c}_F^2} \label{eq:cf}
\end{equation}
The other equation of motion is
\begin{align}
\pd{}{\r} \left[\r^{3-k} \frac{3-k}{2} h_+^{\frac{5-k}{2}} \pd{h_+}{\r} \mathcal{G}\left(y',y,\r\right) + \frac{h_- h_+^{\frac{7-k}{2}}}{2} \frac{1+y'^2}{ \mathcal{G} \left(y',y,\r\right)} \left(\frac{\pd{h_-}{\r} h_+ - h_- \pd{h_+}{\r}}{h_+^2} \right) \right] = \nonumber \\  = \frac{2 y' \r^{3-k} h_+^{\frac{3-k}{2}}}{\mathcal{G}\left(y',y,\r\right)} \label{eq:eomdefect}
\end{align}
with
\begin{equation}
\mathcal{G}\left(y',y,\r\right)=\sqrt{\frac{h^2_-}{h^2_+}\left(1+y'^2\right)+\left(B +2 \p \a' F \right)^2} 
\end{equation}

We find solutions for the embedding function $y(\r)$ by using equation \eq{eq:cf} to eliminate $\left(B +2 \p \a' F \right)$ from \eq{eq:eomdefect} and solving numerically. For Minkowski embeddings, one can show it is not possible to turn on a non constant gauge field $A$ however all chemical potentials are accessible via the asymptotic value of a constant gauge field. This means that the action, which only depends upon the derivative of $A$ does not depend on the chemical potential for Minkowski embeddings, similarly to the Sakai-Sugimoto case. Furthermore, for black hole embeddings there is a constraint on $\tilde{c}_F$ that comes from \eq{eq:cf}. To ensure the reality of $\left(B +2 \p \a' F \right)$ and hence to ensure the chemical potential is purely imaginary we must have
\begin{equation}
\tilde{c}_F < 2^{\frac{3-k}{2}} \r_h^{3-k}
\end{equation}
Since we expect there to be Roberge-Weiss lines at $\m_I \sim \frac{1}{N}$ exploring the behaviour of the theory for small $\tilde{c}_F$ is sufficient.

Once solutions are found we evaluate the renormalised DBI action\footnote{The counterterm that exists for all defect theories is $S_{ct1} = - T N_f \b \mathrm{Vol}(S^{3-k})\mathrm{Vol}(\mathrm{Defect}) \frac{\L^{4-k}}{4-k}$. In the case of $k=2$ defects there is an additional counterterm we will describe below}, dual to the grand canonical ensemble in the gauge theory and observe the transition between the two types of solution. The chemical potential is the asymptotic value of the worldvolume gauge field on the flavour branes, while the expectation value of the phase of the Polyakov loop is similarly related to the NS - NS B field.
\begin{equation}
\a - \b \m_I = \int_{D_2} \left(B + 2 \p \a' F\right) = \tilde{c}_F \b \int_{\r_h}^\infty d\r \frac{h_-}{h_+}\sqrt{ \frac{\left(1+y'^2\right)}{\r^{2(3-k)} h_+^{3-k} -\tilde{c}_F^2}}
\end{equation} 

Consider the theory when the chemical potential is small. Expanding \eq{eq:cf} in powers of $\tilde{c}_F$ and integrating we find 
\begin{equation}
\a - \b \m_I \approx \Psi \tilde{c}_F + \mathcal{O} \left(\tilde{c}_F^3 \right) \label{eq:muexp}
\end{equation}
where
\begin{equation}
\Psi = \b \int_{\r_h}^\infty d\r \sqrt{1+y'^2} \rho ^{k-3} h_-  h_+^{\frac{k}{2}-2}
\end{equation}
$\Psi$ of course depends on the brane embedding but crucially does not depend on $\tilde{c}_F$. We may invert equation \eq{eq:muexp} and substitute into the DBI action \eq{eq:defectaction}. We find
\begin{align}
S_{\mathrm{D}(7-2k)} = T N_f \b \mathrm{Vol}(S^{3-k}) & \mathrm{Vol}(\mathrm{Defect}) \int d\r \sqrt{1+y'^2} \rho ^{3-k} h_- h_+^{\frac{3-k}{2}} \times \nonumber \\ & \times \left(  1 +\frac{  \Psi^2 }{2\rho ^{2(3- k)} h_+^{3-k}}(\a- \b\m_I) ^2 + \mathcal{O} \left(\a- \b\m_I\right)^4  \right)
\end{align}
This shows that when $\a- \b\m_I$ is small the first order correction to the grand canonical ensemble is quadratic in the chemical potential, as expected. 
\subsection{The $k=1$ Case}
A codimension $k=1$ defect means that the glue sector of the theory exists in the full $3+1$ dimensions of the boundary while the fundamental fermionic degrees of freedom live on a $2+1$ dimensional subsurface. For a codimension 1 defect in the limit of zero chemical potential the phase transition between Minkowski and black hole embeddings is of first order as shown in figure \ref{fig:k1s}.
\FIGURE{
\centering
\includegraphics[width=7cm]{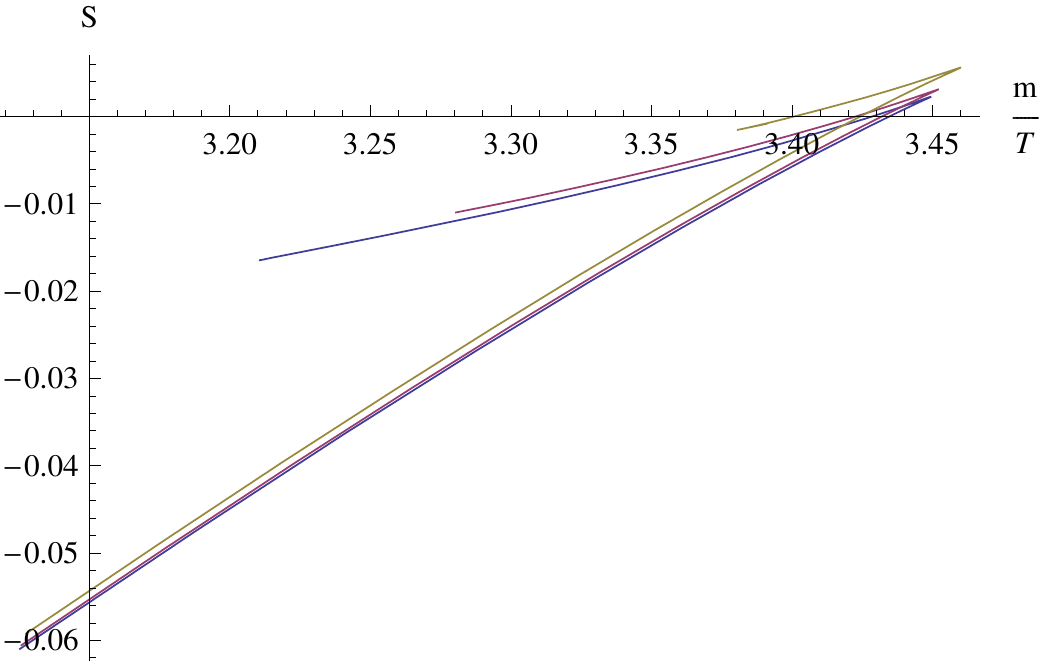}
\includegraphics[width=7cm]{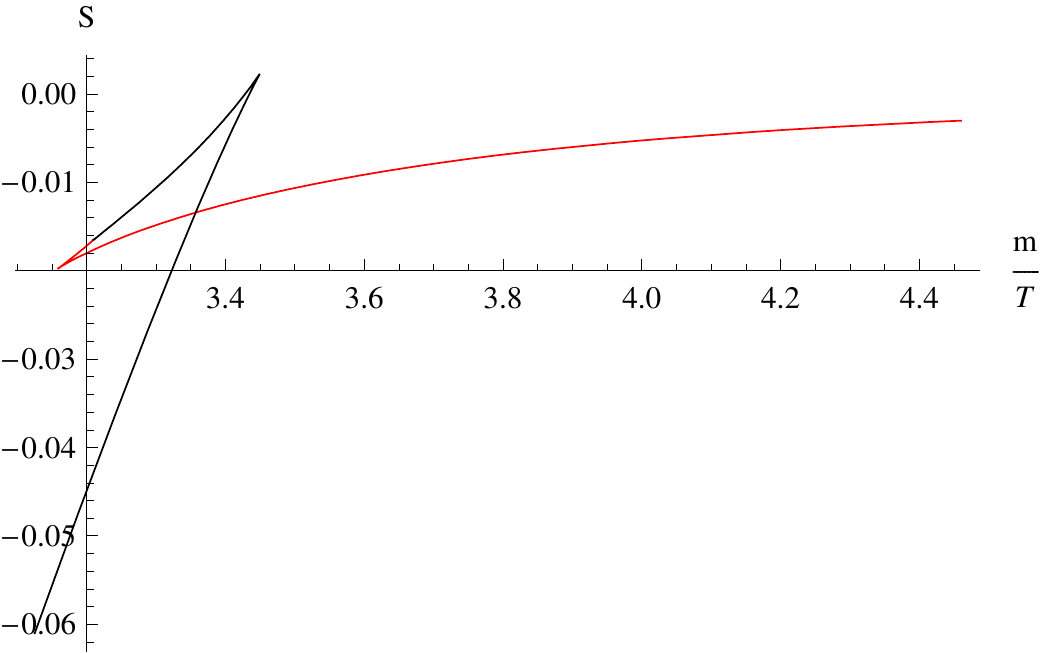}
\caption{For $k=1$: Left; The action as a function of $\frac{m}{T}$ for various $\tilde{c}_F$, for black hole embeddings only. Right; The action as a function of   $\frac{m}{T}$ for $\tilde{c}_F= 0$, both Minkowski and black hole embeddings. \label{fig:k1s}}
}
One can immediately see the characteristic swallowtail shape of the first order phase transition in the grand canonical ensemble for all values of $\tilde{c}_F$ and therefore for all chemical potentials. In \cite{Benincasa:2009be} it was shown that there is a first order phase transition below some critical value of the chemical potential, above which the transition becomes of the second order. In the imaginary chemical potential case the phase transition is always of the first order because the Roberge-Weiss lines appear at order $\frac{1}{N}$ so only the small imaginary chemical potential behaviour of the phase diagram is accessible.

One can compute the phase diagram of the theory by examining the point at which the single line from the Minkowski embeddings intersects the lines that arise from the black hole embeddings. For non zero $\tilde{c}_F$ there is no longer a Minkowski embedding allowed, so the point where the black hole embeddings for some $c_F$ intersect the Minkowski embedding line for $\tilde{c}_F = 0$ is taken to be the position of the phase transition. Recall that one may turn on an arbitrary chemical potential in the Minkowski embeddings by turning on a constant time component of the worldvolume gauge field on flavour branes which does not contribute to the DBI action. Therefore, Non zero chemical potentials are accessible in the Minkowski phase, while non zero $\tilde{c}_F$ is not. 

When we compute the phase diagram we find figure \ref{fig:phasek1}. For small $\m_I$ we the phase transition line is given by 
\begin{equation}
\frac{T}{m} = 0.298 + 0.273 \left(\a-\b\m_I\right)^2 + \mathcal{O} \left(\a -\b \m_I \right)^4
\end{equation}
\FIGURE{
\centering
\includegraphics{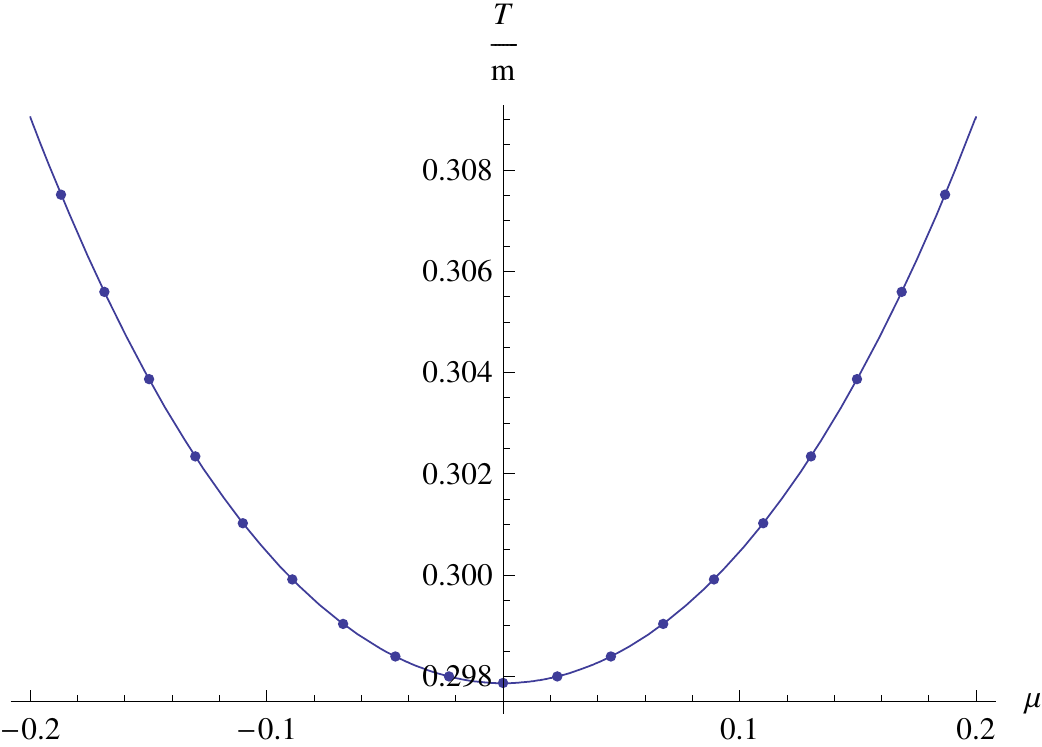}
\caption{The first order meson melting transition line in a codimension 1 defect theory with imaginary chemical potential.}
\label{fig:phasek1}
}
\subsection{The $k=2$ Case}
In this setup the fundamental fermions live in a $1+1$ dimensional subspace of the boundary. The gravity theory is quite different in this case because the embedding function diverges logarithmically at the boundary, which implies by \eq{eq:cf} that the worldvolume gauge field diverges logarithmically at the boundary as well. We still interpret the constant value of the asymptotic embedding function as the quark mass and constant value of the asymptotic worldvolume gauge field as the chemical potential. This means the leading order contribution both of our bulk fields at the boundary is no longer the quark mass and chemical potential in the case of the embedding function and worldvolume gauge field respectively, but the logarithmic term, the coefficient of which is related to the vacuum expectation value of the quark bilinear, and the integral of motion that was defined in \eq{eq:cfdef}. 
\begin{equation}
\lim_{\r \to \infty}y(\r) \approx m + c \log \r + \ldots \quad \lim_{\r \to \infty}A_t(\r) \approx \m + \tilde{c}_F \log \r + \ldots 
\end{equation}
One may show that, as a result of the logarithmic divergences of the embedding function and the worldvolume gauge field, the action requires an additional logarithmic counterterm to renormalise it correctly. The additional counterterm is
\begin{equation}
S_{ct2}= -\frac{1}{2} T N_f \b \mathrm{Vol}(S^{3-k})\mathrm{Vol}(\mathrm{Defect})  \left( c^2 + \tilde{c}_F^2 \right) \log \L
\end{equation}
When we compute the renormalised action as a function of $\frac{m}{T}$ we find a very similar graph to those found before (see figure \ref{fig:k2s}).
\FIGURE{
\centering
\includegraphics[width=7cm]{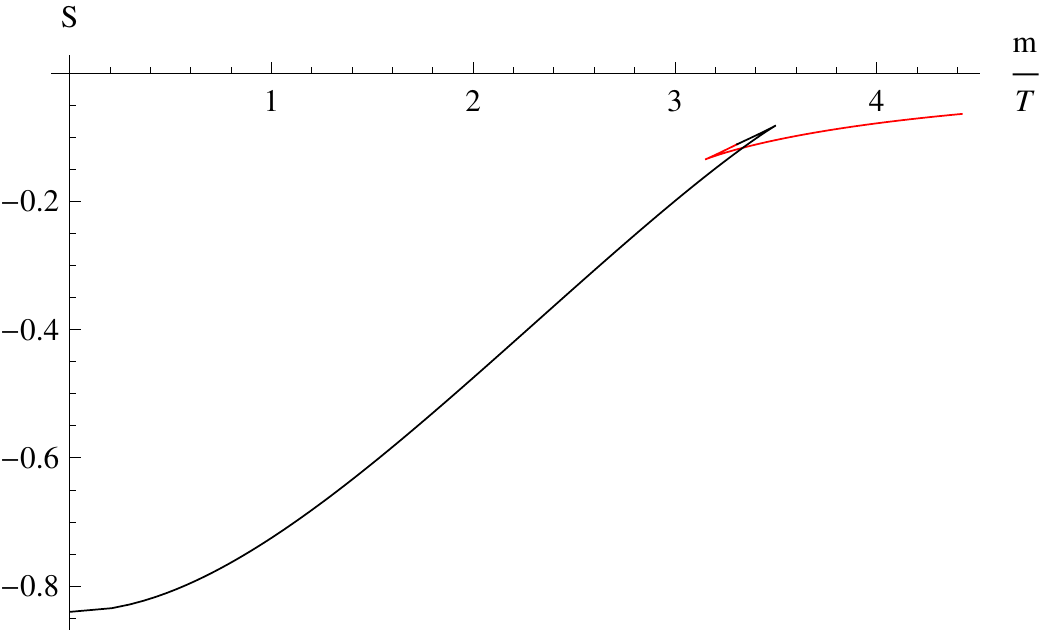}
\includegraphics[width=7cm]{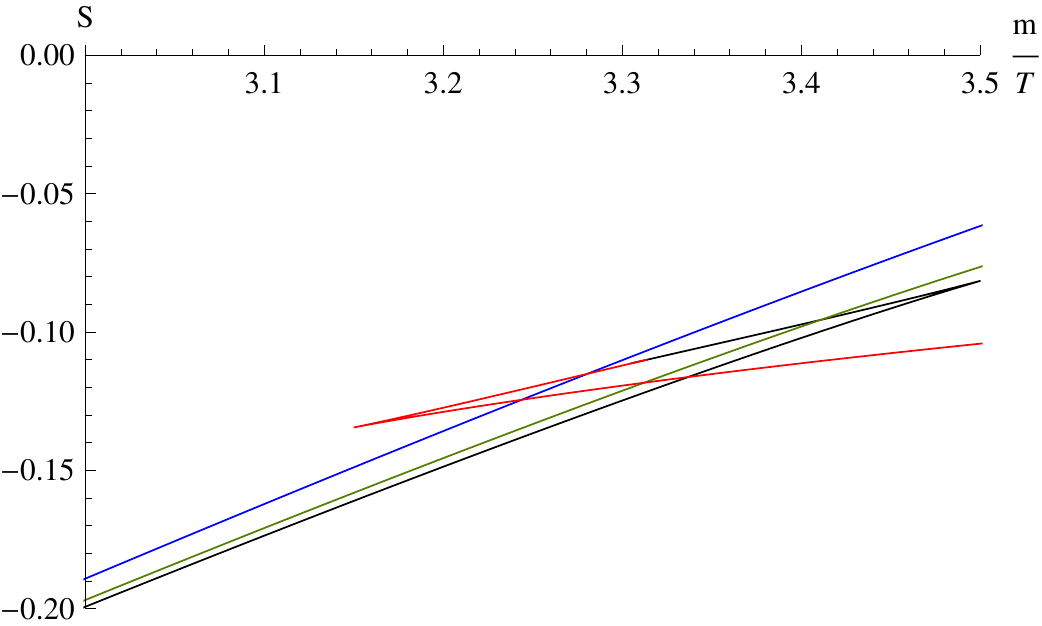}
\caption{The DBI action as a function of $\frac{m}{T}$. On the left is the Minkowski (red) and black hole (black) embedding for $\tilde{c}_F=0$. On the right is a zoomed in version also including black hole embeddings with non zero $\tilde{c}_F$ (green and blue lines). \label{fig:k2s}}
}
The discontinuity in the grand canonical ensemble is clearly first order, and when we compute the position of the first order phase transition we find figure \ref{fig:k2s2}.
\begin{equation}
\frac{T}{m} = 0.2997 + 0.0037 \left(\a - \b \m_I\right)^2 + \mathcal{O}\left( \a - \b \m_I  \right)^4
\end{equation}

\FIGURE{
\centering
\includegraphics{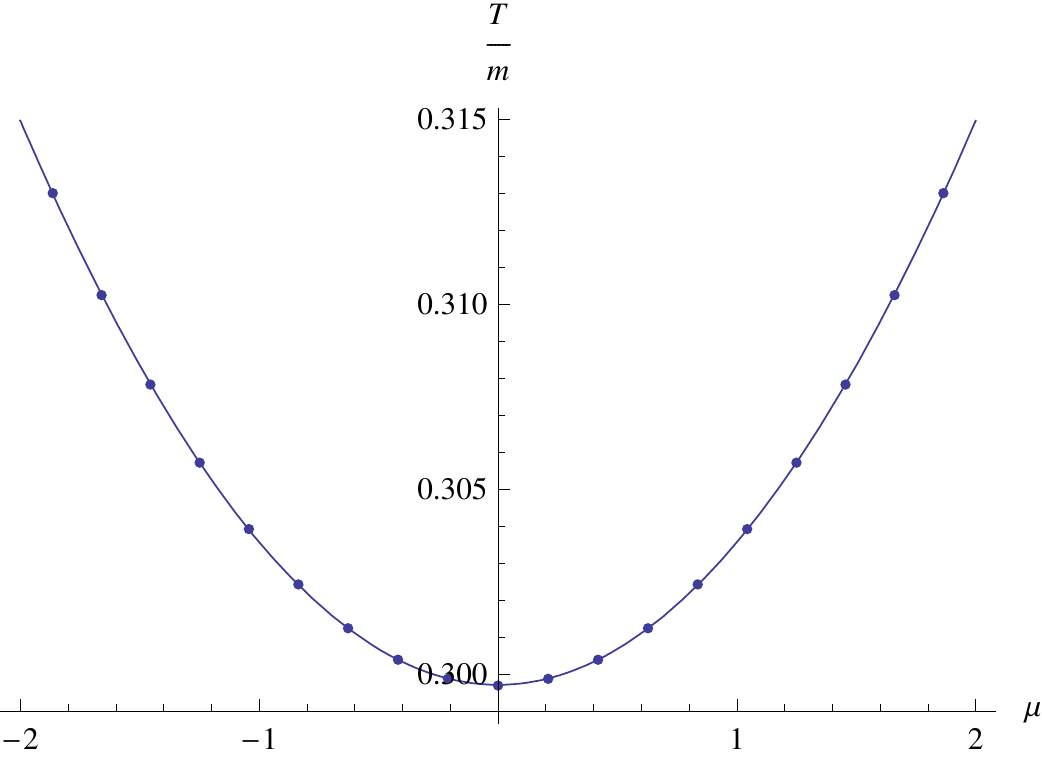}
\caption{The meson melting phase transition line of $k=2$ defect theory in the  $\left(\m_I,\frac{T}{m}\right)$ plane.\label{fig:k2s2}}
}
The phase diagram for codimension 2 systems in $\m^2$ is still not fully known. It was found in \cite{Benincasa:2009be} that there is a second order phase transition meeting the deconfinement transition so there must be a point at some positive $\m^2$ where the phase transition becomes first order, since we have shown the phase diagram to be analytic in $\m^2$ for $\m^2$ close to zero.  
\subsection{Pressure due to Flavours for Zero Quark Mass}
Similarly to the Sakai-Sugimoto case, the pressure is computed by integrating over the bulk directions not on the boundary. 
\begin{align}
P &= - \int d\r d\W_{3-k} \mathcal{L}_\mathrm{DBI} \nonumber\\
P &= - N_f T_k \int d\r d\W_{3-k}\r^{3-k} h_+^{\frac{3-k}{2}} \sqrt{\frac{h^2_-}{h_+}\left(1+y'^2\right)+\left(B +2 \p \a' F \right)^2}
\end{align}
Dimensional analysis tells us that $\left[P\right] = 4-k$ which is expected because the flavours exist only on the defect so the pressure should have the dimension of $- \left[\b\, \mathrm{Vol}\left(\mathrm{Defect} \right) \right] = 4 -k$. Taking the quarks to be massless simplifies things considerably, because this condition implies $y' = y = 0$
\begin{equation}
-N_f T_k V_k \int_{\r_h}^\infty d\r\rho ^{3-k} \left(1+\frac{\sigma_h^4}{\rho ^4}\right)^{\frac{3-k}{2}} \sqrt{\frac{\left(\rho ^4-\s_h^4\right)^2}{\rho ^4 \left(\rho ^4+\s_h^4\right)}+\left(B +2 \p \a' F \right)^2}
\end{equation}
Using the integral of motion \eq{eq:cfdef}
\begin{equation}
P=- N_f T_k V_k\int_{\r_h}^\infty d\r \rho ^{1-k} \left(1+\frac{\s_h^4}{\rho ^4}\right)^{\frac{3-k}{2}} \frac{\left(\rho ^8-\s_h^8\right)}{\sqrt{\left(\rho ^4+\s_h^4\right)^3-\tilde{c}_F^2 \rho ^{6+2 k} \left(1+\frac{\s_h^4}{\rho ^4}\right)^k}}
\end{equation}
We cannot do this integral analytically, so we expand in powers of the density $c_F$ which, when we restore the correct factors of the brane tension and $\a'$ has dimension $3-k$, as we expect. The leading contribution to the pressure is
\begin{equation}
P= -\int_{\r_h}^\infty d\r\frac{N_f T_k V_k \rho ^{1-k} \left(1+\frac{\s_h^4}{\rho ^4}\right)^{\frac{3-k}{2}} \left(\rho ^8-\s_h^8\right)}{\left(\rho ^4+\s_h^4\right)^{\frac{3}{2}}}
\end{equation}
Which, after integration and renormalisation is
\begin{equation}
P = \frac{2^{\frac{3}{2}+\frac{1-k}{2}} N_f T_k V_k \s_h^{4-k}}{4-k}
\end{equation}
We can translate this into field theory quantities using \eq{eq:defectT} and the following relations
\begin{equation}
T_k = \frac{1}{\left(2 \p \right)^{7-2k} l_s^{8-2k}}, \quad \frac{R^4}{l_s^4} = \l
\end{equation}
The volume of a $3-k$ sphere is given by
\begin{equation}
V_k = \frac{\pi ^{\frac{3}{2}-\frac{k}{2}}}{\G\left(\frac{5}{2}-\frac{k}{2}\right)}
\end{equation}
therefore the leading contribution to the pressure is given by
\begin{equation}
P= \frac{2^{2 k-7} N_f \pi ^{-\frac{3-k}{2}+k-3} \lambda ^{\frac{1}{4} (8-2 k)} T^{4-k}}{(4-k) \Gamma\left(1+\frac{3-k}{2}\right)}
\end{equation}
The extra symmetry in this model means the leading order expression for the pressure must be some dimensionless function times $T^{4-k}$, since the temperature is the only dimensionful scale in the problem when the chemical potential is zero. The first order correction when the chemical potential is non zero is
\begin{equation}
-\frac{2^{\frac{1+k}{2}-\frac{5}{2}} c_F^2 \s_h^{k-2}}{\left(2 \p \a' \right)^2 (k-2) N_f  T_k V_k}
\end{equation}
Replacing with field theory quantities as before, the pressure is
\begin{equation}
\frac{4^{2-k} \pi ^{\frac{3}{2}-\frac{k}{2}}  \lambda ^{\frac{k}{2}-1} \Gamma\left(\frac{5}{2}-\frac{k}{2}\right)}{(k-2) N_f} c_F^2 T^{k-2}
\end{equation}
The factor of $k-2$ in the denominator is potentially worrisome, however we shall see that when we rewrite the correction in terms of the chemical potential rather than the density this factor cancels and the expression is well behaved for all values of $k$ we consider. Using the expansion \eq{eq:muexp}, with the additional condition that the quarks are massless and we find
\begin{equation}
c_F \approx \frac{2^{2 k-5} (k-2) }{\Gamma\left(\frac{5}{2}-\frac{k}{2}\right)}N_f \pi ^{\frac{1}{2} (k-3)} T^{2-k} \lambda ^{1-\frac{k}{2}} \mu_I   + \mathcal{O}\left(\m_I^3\right)
\end{equation}
Therefore the first order correction to the pressure is
\begin{equation}
\frac{4^{k-3} (k-2) N_f \pi ^{\frac{1}{2} (k-3)}  \lambda ^{1-\frac{k}{2}} }{\Gamma\left(\frac{5}{2}-\frac{k}{2}\right)}T^{2-k} \mu ^2
\end{equation}
and the pressure for small density is
\begin{equation}
P= \frac{2^{2 k-7} N_f \pi ^{\frac{1}{2} (k-3)} \lambda ^{2-\frac{k}{2}} }{ \Gamma\left(\frac{5}{2}-\frac{k}{2}\right)} T^{4-k}\left( \frac{1}{(4-k)} +\frac{2(k-2)}{\l} \frac{\mu ^2}{T^2} + \mathcal{O}\left(\frac{\m^4}{T^4} \right) \right)
\end{equation}
Note that when $k=2$ the first order correction to the pressure vanishes.
\section{Discussion}
We have computed the phase diagrams of the Sakai-Sugimoto model and codimension $k$ defect theories for $k=1$ and $k=2$ and we have shown that these models are analytic in $\m^2$ when $\m^2$ is small as is to be hoped since we expect these models to be invariant under charge conjugation symmetry, or equivalently the transformation $\m \to -\m$.

In the models we have considered in this paper we find that in the $(\m_I,T)$ phase diagram all transition lines are first order, meaning that the points where the Roberge-Weiss lines meet the phase transition line between low and high temperature phases are triple points. The temperature at which the meson melting transition occurs is proportional to the square of the imaginary chemical potential, which is similar to what was found in \cite{Aarts:2010ky} for the D3/D7 system. 

One may compare the phase diagrams of codimension $k$ defect theories and we note the following, the phase diagram is always of the form 
\begin{equation}
\frac{T}{m} = T_0(k) + T_2 \left(k\right) \left(\a - \b \m_I \right)^2 + \ldots 
\end{equation}
where the coefficients depend on the detail of the defect. We have shown that $T_0$ is always of the same order, but $T_2$ changes significantly between models. Indeed, we note that $T_2 \left(k+1\right)$ is approximately 100 times smaller than $T_2 \left(k\right)$, which means the meson melting transition line is flatter the higher the codimension of the defect. 

When we compute the pressure for small density the defect theories yield what would be expected from dimensional analysis and symmetry arguments - pressure is proportional to $T^{4-k}$ multiplied by a function of $\frac{\m^2}{T^2}$. The Sakai-Sugimoto model on the other hand does not behave as expected, since the pressure goes like $T^7$ at leading order. \\
\textbf{Acknowledgements}
\\ We are very grateful to Gert Aarts, Mark Round, Ross Stanley and particularly Prem Kumar for invaluable discussions. We also thank Ho-Ung Yee for providing comments on initial drafts. JR acknowledges support from the STFC.

\bibliography{rw_refs}{}
\bibliographystyle{JHEP.bst} 
\end{document}